\documentclass{iopart}
\usepackage{xcolor}
\usepackage{bm}
\usepackage{graphicx}
\usepackage{iopams}

\eqnobysec
\usepackage{citesort}

\begin{document}
\title{Discrete time symmetry breaking in quantum circuits: exact solutions
and tunneling}
\author{Feng-Xiao Sun$^{1,2,3,4}$, Qiongyi He$^{1,4,5,6,*}$, Qihuang Gong$^{1,4,5,6}$,
Run Yan Teh$^{3}$, Margaret D. Reid$^{2,3}$ and Peter D. Drummond$^{2,3,\dagger}$}
\address{$^{1}$ State Key Laboratory for Mesoscopic Physics and Collaborative
Innovation Center of Quantum Matter, School of Physics, Peking University,
Beijing 100871, China}
\address{$^{2}$ Institute of Theoretical Atomic, Molecular and Optical Physics
(ITAMP), Harvard University, Cambridge, Massachusetts 02138, USA}
\address{$^{3}$Centre for Quantum and Optical Science, Swinburne University
of Technology, Melbourne 3122, Australia}
\address{$^{4}$ Collaborative Innovation Center of Extreme Optics, Shanxi
University, Taiyuan, Shanxi 030006, China}
\address{$^{5}$ Beijing Academy of Quantum Information Sciences,
Haidian District, Beijing 100193, China}
\address{$^{6}$ Frontiers Science Center for
Nano-optoelectronics, Peking University, Beijing 100871, China}
\ead{$^{*}$ qiongyihe@pku.edu.cn}
\ead{$^{\dagger}$ peterddrummond@gmail.com}
\begin{abstract}
We discuss general properties of discrete time quantum symmetry breaking
in degenerate parametric oscillators. Recent experiments in superconducting
quantum circuit with Josephson junction nonlinearities give rise to
new properties of strong parametric coupling and nonlinearities. Exact
analytic solutions are obtained for the steady-state of this single-mode
case of subharmonic generation. We also obtain analytic solutions
for the tunneling time over which the time symmetry-breaking is lost
above threshold. We find that additional anharmonic terms found in
the superconducting case increase the tunneling rate, and can also
lead to new regimes of tristability as well as bistability. Our analytic
results are confirmed by number state calculations. 
\end{abstract}

\submitto{\NJP}
\maketitle

\section{Introduction}

Quantum time symmetry breaking is a widespread phenomenon in non-equilibrium
quantum optics and superconducting quantum circuits. For quantum clocks
and lasers, time and phase are inter-related. Therefore, quantum time
symmetry breaking occurs in the electromagnetic phase. This is implicit
in the use of coherent states, which have a well-defined phase, to
describe lasers~\cite{glauber1963photon,haken1975cooperative}. Number
state and coherent state descriptions are both complete descriptions
in quantum mechanics. However, coherent state expansions allows one
to recognize more readily that time translational symmetry is broken
through an observation of the phase.

\emph{Discrete} time quantum symmetry breaking takes place in intra-cavity
subharmonic generation, otherwise known as degenerate parametric oscillators~\cite{drummond1980non}.
For quantum optical systems, an exact solution for the steady-state
quantum density matrix~\cite{drummond1981non} is known for the special
case of a single-mode system with no detunings or anharmonicity. From
this, one can calculate quantum tunneling between time phases~\cite{drummond1989quantum}.
Schr\"{o}dinger cats might also seem possible in the steady-state~\cite{wolinsky1988quantum}.
However, this is not the case~\cite{reid1992effect}, although transient
Schr\"{o}dinger cat formation is possible~\cite{reid1992effect,krippner1994transient,munro1995transient}
for strong enough coupling. Tunneling in these systems demonstrates
the existence of long range time order, which has been confirmed in
optical experiments~\cite{nabors1990coherence} that measured extremely
narrow subharmonic line-widths.

Classical spontaneous symmetry breaking is widespread in quantum optics
and related fields, including pattern formation with translational
and cylindrical symmetry breaking~\cite{oppo1994formation}, noncritical
squeezing with rotational symmetry breaking~\cite{navarrete2008noncritically},
and super-solid formation with continuous translational symmetry breaking~\cite{leonard2017supersolid}.
Similar to normal crystals which have a repeating pattern in space,
systems with quantum time symmetry breaking repeat themselves in time.
In the case of spontaneous discrete time-translation symmetry breaking
oscillation takes place at a fraction of the frequency of some periodic
driving force~\cite{else2016floquet}. Here we consider the extent
to which intra-cavity subharmonic generation leads to discrete time
symmetry breaking. This requires an investigation of the time-scale
for restoration of the original symmetry from a symmetry-broken state.

As well as transient macroscopic superpositions and tunneling, subharmonic
generators can also generate space-time ordering~\cite{drummond2005universality}.
These phenomena are often termed as time crystals~\cite{sacha2015modeling,else2016floquet},
since they combine discrete time symmetry breaking with spatial ordering.
Time crystals have also been observed in spin systems~\cite{zhang2017observation,choi2017observation}
and Bose-Einstein condensates~\cite{autti2018observation} experimentally.

As an example of subharmonic generators, degenerate parametric oscillation
has been investigated for a long time in optics~\cite{drummond1980non,drummond1981non,drummond1989quantum,reid1992effect,krippner1994transient,drummond2005universality}.
These devices convert a high-frequency input optical mode into two
equal frequency modes, whose frequency is half of the input mode's,
by means of the parametric nonlinearity. Below threshold, they are
quantum squeezed state generators. These are used to reduce quantum
noise in gravity-wave detectors~\cite{caves1981quantum,pace1993quantum}.
Networks of above threshold parametric devices are now being used
as analog quantum computers for NP-hard optimization~\cite{mcmahon2016fully,inagaki2016coherent}.

The most general case of single mode degenerate parametric oscillation
involves arbitrary detunings, nonlinear losses and anharmonic nonlinearities
in the fundamental mode. This is characteristic, for example, of superconducting
cavity experiments~\cite{leghtas2015confining}. In these investigations,
two superconducting microwave cavities are coupled through a Josephson
junction in a bridge transmon configuration. Mode one, also termed
``the storage'', holds the mode with time symmetry breaking, and
is designed to have minimal single-photon dissipation. Mode two, ``the
readout'', is over-coupled to a transmission line, and it removes
entropy from the storage.

These experiments used Josephson junctions to generate a coupling
that exchanges pairs of photons in the storage with single photons
in the readout. Single-photon dissipation for both modes $a_{1}$
and $a_{2}$ are included. In these systems, as well as other parametric
devices with small mode volumes, there is a large anharmonic nonlinearity,
and possible detunings. These additional effects change the physics,
and modify both the steady-state quantum behavior and the tunneling
rates compared to previous studies.

Quantum tunneling has also been studied in single-mode nonlinear photonic
resonators, both theoretically \cite{drummond1986quantum,drummond1989quantum,Casteels_PRA2016}
and experimentally~\cite{Rodriguez_PRL2017}. In our more general
case there are several parameters to control the tunneling rates,
which were not included in previous work. We study the mean-field
solutions in detail, demonstrating the existence of a universal phase
diagram with monostable, bistable and tristable phases. Including
quantum fluctuations, we obtain the full quantum distribution in all
cases, and focus on the limiting case of pure quantum tunneling at
the midpoint of the hysteresis curve in the bistable region. This
is the most long-lived tunneling region. We show that in this regime
there is an exponentially long tunneling time at large driving. It
is this case where quantum time symmetry breaking has the longest
lifetime before the discrete symmetry is restored.

We first obtain the exact steady-state solution expressed in form
of a complex potential function. The regime where damping exceeds
coupling at the single-photon level is studied in detail. We find
that quantum tunneling occurs in this parameter region, slowly equilibrating
a system with discrete time-symmetry breaking to the steady-state.
The tunneling time is obtained analytically within a novel, complex
potential-barrier approximation in the complex P-representation valid at large particle number. The
results agree with those numerically obtained in the number-state
basis for relatively small photon number. This opens the way to studying
quantum tunneling in multi-mode non-equilibrium systems, as
in the coherent Ising machine \cite{mcmahon2016fully,inagaki2016coherent}.

The paper is organized as follows. In \sref{sec:Time-evolution}
we introduce the Hamiltonian of the two-mode quantum circuit system.
Taking the driving, damping and nonlinearities into account, the master
equation for the time evolution of the system is derived. We use the
generalized P-representation to obtain a conditional Fokker-Planck
equation, and then employ adiabatic elimination of the strongly damped
pump mode to obtain a simpler one-mode Fokker-Planck equation. In
\sref{sec:Mean-field-theory} the mean-field limit is analysed,
showing the existence of stable, bistable and tristable phases.

The quantum steady-state and quantum tunneling are studied in \sref{sec:Quantum-steady-state},
demonstrating that a novel complex manifold method can be used to
calculate tunneling rates. In \sref{sec:Number-state-calculations},
agreement between the analytic results and numerical calculations
using number states is obtained. Finally, \sref{sec:Conclusions}
gives an outlook and our conclusions, with proofs of the complex tunneling
rate method given in an~\ref{sec:Tunneling-rate-for-complex}.

\section{Time-evolution of the system\label{sec:Time-evolution}}

\subsection{General model Hamiltonian}

Quantum optical and quantum circuit physics are closely related. The
principal difference is that quantum circuits operate at much lower
temperatures, and at microwave rather than optical frequencies. To
treat both cases, we consider a general model for nonlinear interactions
and damping of two coupled bosonic modes of an open system. We include
driving, damping and both coherent and dissipative nonlinearities
in the degenerate parametric oscillator.

We suppose that $a_{k},a_{k}^{\dagger}$ are the $k$-th mode annihilation
and creation operators of modes at two different frequencies $\omega_{k}$
in coupled resonant cavities. There is an overall quantum Hamiltonian
given by 
\begin{equation}
H=\hbar\sum_{n=1}^{6}\sum_{k=1}^{2}H_{k}^{(n)},
\end{equation}
where $H_{k}=\sum_{n}H_{k}^{(n)}$ is the Hamiltonian for the $k$-th
mode, with driving, damping, and nonlinear terms. We define $H^{(n)}=\sum_{k}H_{k}^{(n)}$
as the sum over modes for the $n-th$ type of interaction. The detailed
structure of these terms is as follows: 
\begin{eqnarray}
H_{k}^{(1)} & =\left[\hat{\Gamma}_{k}a_{k}^{\dagger}+h.c.\right]+H_{k}^{R},\nonumber \\
H_{k}^{(2)} & =\left[\hat{\Gamma}_{k}^{(2)}a_{k}^{\dagger2}+h.c.\right]+H_{k}^{R2},\nonumber \\
H_{k}^{(3)} & =\omega_{k}a_{k}^{\dagger}a_{k},\nonumber \\
H_{k}^{(4)} & =i\mathcal{E}_{k}a_{k}^{\dagger}e^{-ik\omega t}+h.c.,\nonumber \\
H_{1}^{(5)} & =\frac{\chi}{2}a_{1}^{\dagger2}a_{1}^{2},\nonumber \\
H_{1}^{(6)} & =i\frac{\kappa}{2}a_{2}a_{1}^{\dagger2}+h.c.\label{eq:hamiltonian}
\end{eqnarray}
Here $\hat{\Gamma}_{k}$ are the external reservoir coupling operators
with reservoir Hamiltonians $H_{k}^{R}$, and $\mathcal{E}_{k}$ are
the envelope amplitudes of external coherent driving fields at angular
frequency $k\omega$ for the $k$-th mode. The nonlinear parameters
are $\kappa$ for subharmonic generation and $\chi$ for anharmonicity.
We suppose $\omega_{2}\simeq2\omega_{1}\simeq2\omega$, so the system
can be externally driven simultaneously at fundamental and subharmonic
frequencies, although we include detunings as well.

In many typical experiments~\cite{leghtas2015confining}, the inputs
are in the higher frequency mode, $a_{2}$. Thus, the subharmonic
driving field is zero for such experiments. We will keep this term
in the general Hamiltonian and exact solutions, for completeness,
but set it to zero in the tunneling calculations below.

\subsection{Master Equation}

The system Hamiltonian $H_{rev}$ is defined as the reversible part
of the Hamiltonian without reservoir couplings, so that: 
\begin{equation}
H_{rev}=\hbar\sum_{n=3}^{6}H^{(n)}.
\end{equation}
The rotating frame system Hamiltonian $H^{S}$ is obtained by subtracting
the driving frequency terms. This is used to give a picture such that
only slowly varying behavior is retained in the state equation, while
the operators evolve at their respective driving frequencies. Therefore,
we define an interaction picture such that: 
\begin{equation}
a_{k}^{\dagger}\left(t\right)=a_{k}^{\dagger}(0)e^{-ik\omega t},
\end{equation}
and evolve the density matrix using the subtracted Hamiltonian, $H^{SR}$.
From now on we let $a_{k}^{\dagger}=a_{k}^{\dagger}(0)$, and define:
\begin{equation}
H^{SR}=H_{rev}-\sum_{k}k\hbar\omega a_{k}^{\dagger}a_{k}.
\end{equation}
This transformation has the effect of changing the mode frequencies
in the Hamiltonian to relative detunings, so that $\omega_{k}\rightarrow\Delta_{k}=\omega_{k}-k\omega$.
The resulting master equation~\cite{gardiner2004quantum,drummond2014quantum}
for the quantum density matrix $\rho$, on tracing over all the reservoirs
in the Markovian approximation, is: 
\begin{equation}
\dot{\rho}=-\frac{i}{\hbar}\left[H^{SR},\rho\right]+\sum_{k,j>0}\frac{1}{j}\gamma_{k}^{(j)}\mathcal{L}_{k}^{(j)}\left[\rho\right]\,.
\end{equation}
Here $\gamma_{k}^{(j)}$ are the linear and nonlinear amplitude relaxation
rates for $j$-boson relaxation in the $k$-th mode, and $n_{k}^{(j)}$
are the corresponding thermal occupations of the reservoirs. These
are $n_{k}^{(j)}=1/\left[\exp\left(j\hbar\omega_{k}/k_{B}T\right)-1\right]$,
where $j\omega_{k}$ is the resonant frequency of the $k$-th mode
reservoir for the $j$-photon damping process, and $k,j=1,2$. Note
that either resonant mode could have single-photon and/or two-photon
losses~\cite{chaturvedi1977two}, but for simplicity we only include
nonlinear losses for $k=1$. Higher order decoherence is also possible~\cite{opanchuk2013functional},
but not treated here.

We set $\hat{O}=\hat{a}_{k}^{j}$ to describe general $j$-photon
damping in the $k$-th mode, giving the the master equation super-operator
for decoherence as 
\begin{equation}
\mathcal{L}_{k}^{(j)}\left[\rho\right]=2\hat{O}\rho\hat{O}^{\dagger}-\rho\hat{O}^{\dagger}\hat{O}-\hat{O}^{\dagger}\hat{O}\rho+2n_{k}^{(j)}\left[\left[\hat{O},\rho\right],\hat{O}^{\dagger}\right].
\end{equation}
We will put $n_{2}^{(1)}=n_{k}^{(2)}=0$, for simplicity, since these
reservoirs have at least twice the frequency of the fundamental reservoirs,
and will have lower thermal occupations, which we neglect. The energy
relaxation rate in each mode for single-particle decay is $2\gamma_{k}^{(1)}$.
In the exact solutions presented in later sections, we also assume
that $n_{1}^{(1)}=0$. This allows us to investigate the important
exact quantum tunneling solutions in the low-temperature limit, although
finite temperature effects in the fundamental reservoirs are included
in the next section for completeness. We include thermal occupation
for single-photon processes at this stage, to show how this alters
the resulting equations in high temperature cases where such effects
are important.

\subsection{Fokker-Planck equation\label{sec:Fokker-Planck-and-stochastic}}

We now introduce the generalized P-representation developed in the
reference~\cite{drummond1980generalised}. This expands the quantum
density matrix in terms of a complete operator basis $\hat{\Lambda}\left(\bm{\alpha},\bm{\alpha}^{+}\right)$,
and a P-representation $P\left(\bm{\alpha},\bm{\alpha}^{+}\right)$
so that: 
\begin{equation}
\hat{\rho}=\int\int d\mu\left(\bm{\alpha},\bm{\alpha}^{+}\right)P\left(\bm{\alpha},\bm{\alpha}^{+}\right)\hat{\Lambda}\left(\bm{\alpha},\bm{\alpha}^{+}\right).
\end{equation}
The operator basis uses off-diagonal coherent state projection operators,
and has the form, $\hat{\Lambda}\left(\bm{\alpha},\bm{\alpha}^{+}\right)=\left|\bm{\alpha}\right\rangle \left\langle \bm{\alpha}^{+*}\right|/\left\langle \bm{\alpha}^{+*}\right|\left.\bm{\alpha}\right\rangle $,
where $\left|\bm{\alpha}\right\rangle $ is a two-mode coherent state.
This includes a normalizing factor so that the P-distribution integrates
to unity. The integration measures $d\mu$ can be chosen as either
a volume measure on a full eight-dimensional complex space over the
combined vector $\vec{\alpha}=\left(\bm{\alpha},\bm{\alpha}^{+}\right)$,
or as a contour integral on a complex manifold, which is explained
in detail below. The advantage of this approach is that the resulting
Fokker-Planck equation is exact without truncation. This is not the
case, for example, with the Wigner representation. In the quantum-dominated
regime with a strongly damped high-frequency mode, we show that there
is an exact solution for the equilibrium steady-state of the resulting
single-mode Fokker-Planck equation.

All normally ordered quantum correlation functions are moments of
the distribution, since: 
\begin{equation}
\left\langle a_{j}^{\dagger}\ldots a_{k}\right\rangle =\int\int d\mu\left(\bm{\alpha},\bm{\alpha}^{+}\right)P\left(\bm{\alpha},\bm{\alpha}^{+}\right)\left[\alpha_{j}^{+}\ldots\alpha_{k}\right].\label{eq:nonclassical correlation-1}
\end{equation}
Using standard operator identities, the resulting Fokker-Planck equation
has the form that: 
\begin{equation}
\frac{\partial P}{\partial t}=\sum_{n=1}^{6}\sum_{k=1}^{2}\mathcal{D}_{k}^{(n)}P,
\end{equation}
where we introduce the notation that $\mathcal{D}_{k}^{(n)}$ is a
differential operator acting on the P-function. The derivative operators
are: $\partial_{k}\equiv\partial/\partial\alpha_{k}$, $\partial_{k}^{+}\equiv\partial/\partial\alpha_{k}^{+}$,
and the individual terms involved that correspond to each Hamiltonian
coupling are: 
\begin{eqnarray}
\mathcal{D}_{k}^{(1)} & =\gamma_{k}^{(1)}\left[\left(\partial_{k}\alpha_{k}+h.c\right)+n_{k}^{(1)}\partial_{k}\partial_{k}^{+}\right],\nonumber \\
\mathcal{D}_{1}^{(2)} & =\gamma_{1}^{(2)}\left[\partial_{1}\alpha_{1}^{+}\alpha_{1}^{2}-\frac{1}{2}\partial_{1}^{2}\alpha_{1}^{2}\right]+h.c.,\nonumber \\
\mathcal{D}_{k}^{(3)} & =i\partial_{k}\Delta_{k}\alpha_{k}+h.c.,\nonumber \\
\mathcal{D}_{k}^{(4)} & =-\partial_{k}\mathcal{E}_{k}+h.c.,\\
\mathcal{D}_{1}^{(5)} & =i\chi\left[\partial_{1}\alpha_{1}^{+}\alpha_{1}^{2}-\frac{1}{2}\partial_{1}^{2}\alpha_{1}^{2}\right]+h.c.,\nonumber \\
\mathcal{D}_{1}^{(6)} & =\kappa\left[-\partial_{1}\alpha_{1}^{+}\alpha_{2}+\frac{1}{2}\partial_{2}\alpha_{1}^{2}+\frac{1}{2}\partial_{1}^{2}\alpha_{2}\right]+h.c.\nonumber 
\end{eqnarray}

Here h.c. denotes a term generated from the hermitian conjugate operator
identities, in which coefficients are conjugated, and $\alpha_{k}\rightarrow\alpha_{k}^{+}$.
Due to the freedom to make phase rotations in defining the mode operators,
we can define $\mathcal{E}_{2},\kappa$ to be real parameters without
loss of generality.

By combining terms, we obtain the following Fokker-Planck equation,
where $\mu=1,\ldots4$, $\vec{\partial}=\left(\partial_{\bm{\alpha}},\partial_{\bm{\alpha}^{+}}\right)$,
and an Einstein summation convention is used to sum over repeated
indices:

\begin{equation}
\frac{\partial P}{\partial t}=\left[-\partial_{\mu}A_{\mu}+\frac{1}{2}\partial_{\mu}\partial_{\nu}D_{\mu\nu}\right]P.
\end{equation}

In this form it is convenient to introduce complex single particle
decay rates $\gamma_{k}$ and two-particle decay rates $g_{1}$, so
that these parameters can be combined into complex rate terms: 
\begin{equation}
\gamma_{k}=\gamma_{k}^{(1)}+i\Delta_{k},\quad g_{1}=\gamma_{1}^{(2)}+i\chi.
\end{equation}
With these definitions, the combined deterministic or drift term becomes:
\begin{equation}
\vec{A}\left(\vec{\alpha}\right)=\left[\begin{array}{c}
\mathcal{E}_{1}-\gamma_{1}\alpha_{1}-g_{1}\alpha_{1}^{+}\alpha_{1}^{2}+\kappa\alpha_{1}^{+}\alpha_{2}\\
\mathcal{E}_{1}^{*}-\gamma_{1}^{*}\alpha_{1}^{+}-g_{1}^{*}\alpha_{1}\alpha_{1}^{+2}+\kappa\alpha_{1}\alpha_{2}^{+}\\
\mathcal{E}_{2}-\gamma_{2}\alpha_{2}-\kappa\alpha_{1}^{2}/2\\
\mathcal{E}_{2}-\gamma_{2}^{*}\alpha_{2}^{+}-\kappa\alpha_{1}^{+2}/2
\end{array}\right].
\end{equation}
The corresponding combined diffusion coefficient is then: 
\begin{equation}
\underline{\underline{D}}\left(\vec{\alpha}\right)=\left[\begin{array}{cc}
\underline{D}_{1}\\
 & \underline{0}
\end{array}\right],
\end{equation}
where the individual mode diffusion sub-matrices are: 
\begin{eqnarray}
\underline{D}_{1} & =\left[\begin{array}{cc}
\kappa\alpha_{2}-g_{1}\alpha_{1}^{2} & \Gamma_{1}\\
\Gamma_{1} & \kappa\alpha_{2}^{+}-g_{1}^{*}\alpha_{1}^{+2}
\end{array}\right].
\end{eqnarray}
while $\Gamma_{1}\equiv2\gamma_{1}^{(1)}n_{1}^{(1)}$ is the thermal
noise coefficient.

\subsection{Stochastic Equations\label{subsec:Stochastic-Equations}}

If all modes have strong enough damping, so that all boundary terms
vanish in the Fokker-Planck equation, there is a corresponding stochastic
equation for the positive P-representation \cite{drummond1980generalised},
which can be written in a combined vector form as: 
\begin{equation}
\frac{d\vec{\alpha}}{dt}=\vec{A}\left(\vec{\alpha}\right)+\underline{\underline{B}}\left(\vec{\alpha}\right)\vec{\zeta}\left(t\right),\label{eq:SDE}
\end{equation}
where the Gaussian noise term $\zeta$ has a vanishing mean, and the
only nonzero correlations are: 
\begin{equation}
\left\langle \zeta_{i}\left(t\right)\zeta_{j}\left(t'\right)\right\rangle =\delta_{ij}\delta\left(t-t'\right).\label{eq:noise}
\end{equation}
The corresponding combined stochastic coefficient is then: 
\begin{equation}
\underline{\underline{B}}\left(\vec{\alpha}\right)=\left[\begin{array}{cc}
\underline{B}_{1}\\
 & \underline{0}
\end{array}\right],
\end{equation}
where the individual mode noise sub-matrices are: 
\begin{eqnarray}
\underline{B}_{1} & =\left[\begin{array}{cc}
\kappa\alpha_{2}-g_{1}\alpha_{1}^{2} & \Gamma_{1}\\
\Gamma_{1} & \kappa^{*}\alpha_{2}^{+}-g_{1}^{*}\alpha_{1}^{+2}
\end{array}\right]^{1/2}.
\end{eqnarray}

In the next section, we will focus on the steady-state solutions in
the zero temperature limit, in order to understand the steady-state
properties of maximal quantum coherence. Although these stochastic
equations are useful when the damping rates of both modes are large
compared to the nonlinearities, they have no known analytic solutions.
In addition, these stochastic equations can have boundary terms at
strong coupling, leading to instabilities. For this reason, we turn
next to a hybrid method. This allows us to derive a solution for the
complex P-representation of the subharmonic mode, which allows us
to analytically calculate the tunneling rate.

\subsection{Hybrid representation}

We will consider the case where the second harmonic mode is strongly
damped, and the first harmonic mode is not, as in many experiments.
This will be treated in a hybrid measure, where the second harmonic
mode is treated stochastically, while the first harmonic is expanded
on a complex manifold. In this case, we extend methods used previously
through defining a conditional $P_{2}$ distribution for the second
harmonic mode that depends on the amplitude of the first mode, so
that: 
\begin{equation}
\hat{\rho}=\int_{\mathcal{C}}d\vec{\alpha_{1}}P_{1}\left(\vec{\alpha}_{1}\right)\int d\vec{\alpha_{2}}P_{2}\left(\vec{\alpha}_{2}\left|\vec{\alpha}_{1}\right.\right)\hat{\Lambda}\left(\bm{\alpha},\bm{\alpha}^{+}\right).
\end{equation}
Since this $\vec{\alpha}_{2}$ mode is strongly damped, it can be
readily solved on the relevant short time-scales. This is equivalent
to a standard characteristic function solution of a first order partial
differential equation: 
\begin{equation}
\dot{\alpha}_{2}=\mathcal{E}_{2}-\gamma_{2}\alpha_{2}-\kappa\alpha_{1}^{2}/2.
\end{equation}
In the limit of $\gamma_{2}^{(1)}\gg\gamma_{1}^{(1)}$, the second
harmonic mode is rapidly damped to a deterministic solution, 
\begin{equation}
\alpha_{2}^{(a)}=\frac{\mathcal{E}_{2}-\kappa\alpha_{1}^{2}/2}{\gamma_{2}}.\label{eq:adiabatic_a2}
\end{equation}
There is a similar equation for $\alpha_{2}^{+}$, which means that
in the adiabatic limit 
\begin{equation}
P_{2}\left(\vec{\alpha}_{2}\left|\vec{\alpha}_{1}\right.\right)=\delta\left(\vec{\alpha}_{2}-\vec{\alpha}_{2}^{(a)}\right).
\end{equation}

For simplicity, we assume that in this strong damping limit the corresponding
detuning $\Delta_{2}$ is negligible, and therefore $\gamma_{2}$
is treated as a real parameter. The details of adiabatic elimination
in the full quantum theory are treated in the next section.

\subsection{Quantum adiabatic elimination\label{sec:Adiabatic-Elimination}}

We will treat the quantum effects in the adiabatic limit, but with
quantum noise included. We now introduce a reduced P-representation
obtained by tracing over the second-harmonic mode, so that $\hat{\rho}_{1}=\Tr_{2}\left(\hat{\rho}\right)$.
If we expand the quantum density matrix in terms of a single-mode
operator basis $\hat{\Lambda}_{1}\left(\vec{\alpha}\right)$, and
a single-mode P-representation $P\left(\vec{\alpha}\right)$, where
$\vec{\alpha}=\vec{\alpha}_{1}$, one then obtains: 
\begin{equation}
\hat{\rho}_{1}=\int\int d\mu\left(\vec{\alpha}\right)P_{1}\left(\vec{\alpha}\right)\hat{\Lambda}_{1}\left(\vec{\alpha}\right).
\end{equation}
The operator basis uses coherent state projection operators, as before,
but with a simpler form, $\hat{\Lambda}_{1}\left(\vec{\alpha}\right)\equiv\left|\alpha\right\rangle \left\langle \alpha^{+*}\right|/\left\langle \alpha^{+*}\right|\left.\alpha\right\rangle $.
All normally ordered single-mode quantum correlation functions are
moments of the distribution, since: 
\begin{equation}
\left\langle a^{\dagger}\ldots a\right\rangle =\int\int d\mu\left(\vec{\alpha}\right)P_{1}\left(\vec{\alpha}\right)\left[\alpha^{+}\ldots\alpha\right].\label{eq:nonclassical correlation}
\end{equation}
With this approach, the elimination of the $\alpha_{2}$ amplitude
results in a single-mode Fokker-Planck equation, 
\begin{equation}
\frac{\partial P_{1}}{\partial t}=\left\{ \frac{\partial}{\partial\alpha}\left[\gamma\alpha-\mathcal{E}_{1}-\bar{\mathcal{E}}\left(\alpha\right)\alpha^{+}\right]+\frac{1}{2}\frac{\partial^{2}}{\partial\alpha^{2}}\bar{\mathcal{E}}\left(\alpha\right)+hc\right\} P_{1}.\label{eq:FPE}
\end{equation}
Here $\gamma\equiv\gamma_{1}$ and the combined effective nonlinear
loss is: 
\begin{equation}
\gamma^{(2)}=\gamma_{1}^{(2)}+\frac{\kappa^{2}}{2\gamma_{2}}.
\end{equation}
This leads to a combined nonlinear coefficient $g$, where: 
\begin{equation}
g=g_{1}+\kappa^{2}/2\gamma_{2}=\gamma^{(2)}+i\chi.
\end{equation}
We have also defined 
\begin{equation}
\bar{\mathcal{E}}\left(\alpha\right)\equiv\frac{\kappa}{\gamma_{2}}\left[\mathcal{E}_{2}-\kappa\alpha^{2}/2\right]-g_{1}\alpha^{2}=\mathcal{E}-g\alpha^{2},
\end{equation}
 with $\mathcal{E}\equiv\kappa\mathcal{E}_{2}/\gamma_{2}$. Physically,
$\bar{\mathcal{E}}\left(\alpha\right)$, together with
its conjugate, $\bar{\mathcal{E}}^{*}\left(\alpha^{+}\right)$,
are input fields that include depletion. The notation $hc$ indicates
hermitian conjugate terms obtained by the replacement of $\alpha\rightarrow\alpha^{+}$,
and the conjugation of all complex parameters. In the present case
that $\Delta_{2}=0$, we have a real $\gamma_{2}$ and a real $\mathcal{E}$,
but $g$ is still generally complex because of the nonzero anharmonic
nonlinearity.

It is simpler for the detailed analysis of this problem to use dimensionless
parameters, which we define as follows: 
\begin{eqnarray}
\epsilon & = & \mathcal{E}/g,\nonumber \\
n & = & \left|\epsilon\right|,\nonumber \\
c & = & \gamma/\left(gn\right),\nonumber \\
\tau & = & \mathcal{E}t,\nonumber \\
d & = & \frac{\left|g\right|\mathcal{E}_{1}}{g\mathcal{E}\sqrt{\epsilon}},\nonumber \\
\beta & = & \alpha/\sqrt{\epsilon}\,.\label{eq:parameter}
\end{eqnarray}

We also introduce a relative phase, $e^{i\theta}=g/|g|=n/\epsilon$.
This gives a Fokker-Planck equation in a more universal form, as:
\begin{eqnarray}
\frac{\partial P_{1}\left(\vec{\beta}\right)}{\partial\tau} & = & e^{i\theta}\left\{ \frac{\partial}{\partial\beta}\left[c\beta-d-\left(1-\beta^{2}\right)\beta^{+}\right]+\right.\nonumber \\
 &  & +\left.\frac{1}{2n}\frac{\partial^{2}}{\partial\beta^{2}}\left(1-\beta^{2}\right)+h.c.\right\} P_{1}\left(\vec{\beta}\right).\label{FPE_scaled}
\end{eqnarray}

Physically, time is scaled relative to the two-photon driving rate,
while $c$ is a complex dimensionless single-photon loss and detuning.
The important scaling parameter $n$ is the photon number at which
saturation of the mode occupation occurs due to the nonlinear losses.

\subsection{Equivalent Hamiltonian}

After adiabatically eliminating the second harmonic mode~\cite{drummond1981non,leghtas2015confining},
we obtained a new Fokker-Planck equation, as described above. This
corresponds exactly to the dynamics for the fundamental quantum system
with $a\equiv a_{1}$, governed by the adiabatic Hamiltonian, 
\begin{equation}
\frac{H_{A}}{\hbar}=\Delta_{1}a^{\dagger}a+i\left[\mathcal{E}_{1}a^{\dagger}+\frac{\mathcal{E}}{2}a^{\dagger2}-h.c.\right]+\frac{\chi}{2}a^{\dagger2}a^{2},\label{Hamiltonian}
\end{equation}
together with a single-photon loss $\gamma^{(1)}=\gamma_{1}^{(1)}$
and an effective two-photon loss $\gamma^{(2)}$ which is defined
above.

The master equation of the density matrix $\rho_{1}$ that is equivalent
to the new Fokker-Planck equation given above is then obtained as
\begin{eqnarray}
\frac{\partial}{\partial t}\rho_{1} & = & \frac{1}{i\hbar}[H_{A},\rho_{1}]+\gamma^{(1)}(2a\rho_{1}a^{\dagger}-a^{\dagger}a\rho_{1}-\rho_{1}a^{\dagger}a)\nonumber \\
 &  & +\frac{\gamma^{(2)}}{2}(2a^{2}\rho_{1}a^{\dagger2}-a^{\dagger2}a^{2}\rho_{1}-\rho_{1}a^{\dagger2}a^{2}),\label{eq:master_eq}
\end{eqnarray}
where we have used a zero-temperature limit. Similar systems have
been studied before where quantum squeezing and bifurcation have been
found~\cite{kryuchkyan1996exact,meaney2014quantum,bartolo2016exact,elliott2016applications},
but here we focus on the issue of quantum tunneling.

Equation~(\ref{eq:master_eq}) can be solved numerically using a
number-state expansion provided the photon number is not too large.
However, to gain more insight from analytic results, it is also possible
to use the P-representation directly, given the single-mode Fokker-Planck
equation~(\ref{eq:FPE}). In the following sections, we will employ
both methods.

\section{Mean field theory predictions \label{sec:Mean-field-theory}}

Before carrying out the full quantum theory calculations, we obtain
the predictions of mean field theory for the degenerate parametric
oscillator, where the system is described by the equation (\ref{eq:SDE})
but without fluctuations. The resulting deterministic equations are
a good approximation when the system has negligible quantum and thermal
noise. Since the noise terms are generated by nonlinearities, this
implies that the nonlinearity at the single photon level is much smaller
than the damping. From the dimensionless Fokker-Planck equation, equation
(\ref{FPE_scaled}), the mean-field limit corresponds to $n\rightarrow\infty$.

Earlier work on the mean field theory in single-mode driven systems
included a degenerate parametric oscillator without anharmonicity
(i.e. $\chi=0$) \cite{drummond1980non}, and a purely anharmonic
system without parametric terms \cite{drummond1980quantum}. The steady
state solutions for the deterministic equations were found and their
stability was studied by considering the behavior of small perturbations
around these steady state solutions. This work showed that the subharmonic
mode has a steady state with zero mean amplitude when the pump amplitude
is below a certain critical value. Above this critical value, the
zero mean subharmonic mode amplitude is no longer a stable steady
state solution. Rather, the mode has a bistable solution where the
two steady states have an equal amplitude but with a phase difference
of $\pi$. We now show that the situation when there is both anharmonic
and detuning terms added is more complex than this.

\subsection{Mean field dynamics}

In the mean field case, one sets $\alpha_{k}^{+}=\alpha_{k}^{*}$,
and treats the dynamical equation of

\begin{equation}
\frac{d\bm{\alpha}}{dt}=\bm{A}\left(\bm{\alpha}\right),\label{eq:SDE-1-1-1}
\end{equation}
where the drift terms are

\begin{equation}
\bm{A}\left(\bm{\alpha}\right)=\left[\begin{array}{c}
\mathcal{E}_{1}-\gamma_{1}\alpha_{1}-g_{1}\alpha_{1}^{*}\alpha_{1}^{2}+\kappa\alpha_{1}^{*}\alpha_{2}\\
\mathcal{E}_{2}-\gamma_{2}\alpha_{2}-\kappa\alpha_{1}^{2}/2
\end{array}\right].
\end{equation}
Here the single photon damping rates $\gamma_{1}$ and $\gamma_{2}$
are generally complex. For $\gamma_{2}^{(1)}\gg\gamma_{1}^{(1)}$,
one can adiabatically eliminating the mode $\alpha_{2}$, just as
in the Fokker-Planck approach of equation (\ref{eq:adiabatic_a2}).

Next, we set $\mathcal{E}_{1}=0$ in the mean field calculation, to
correspond to the bistable situation of most interest here. The mean-field
equation for the mode amplitude $\alpha$ is then 
\begin{equation}
\dot{\alpha}=-\left[\gamma+g\left|\alpha\right|^{2}\right]\alpha+\mathcal{E}\alpha^{*}.
\end{equation}

To simplify this, we introduce the dimensionless parameters of (\ref{eq:parameter}),
which normalize the damping by the driving. These equations can also
be obtained directly from the scaled Fokker-Planck equation (\ref{FPE_scaled}).
On scaling, the second-order derivative terms can be neglected in
the mean-field limit of $n\gg1$, leading to the mean-field dynamical
equations:
\begin{equation}
\frac{\partial\beta}{\partial\tau}=e^{i\theta}\left[\left(1-\beta^{2}\right)\beta^{*}-c\beta\right].
\end{equation}

\subsection{Mean-field stationary states}

To obtain the mean-field stationary values, we set $\dot{\beta}=\dot{\beta}^{*}=0$,
so that
\begin{eqnarray}
\beta^{*} & = & \left[\left|\beta\right|^{2}+c\right]\beta\label{eq:steady-state}\\
 & = & \left|\left|\beta\right|^{2}+c\right|^{2}\beta^{*}.\nonumber 
\end{eqnarray}
This has the solutions $\beta=0$, together with the solutions of
the quadratic 
\begin{equation}
\left|c\right|^{2}-1+\left(c+c^{*}\right)\left|\beta\right|^{2}+\left|\beta\right|^{4}=0\,.
\end{equation}

The quadratic has two roots for the intra-cavity intensity:

\begin{equation}
\left|\beta\right|^{2}=-\Re\left(c\right)\pm\Pi(c),\label{eq:intensity}
\end{equation}
where:
\begin{equation}
\Pi(c)\equiv\sqrt{1-\Im(c)^{2}}\,.
\end{equation}
Since this is an intensity, a negative solution or complex solution
is not possible. A real solution clearly requires $|\Im(c)|\le1$.
After calculating the corresponding amplitude in each case of a non-negative
intensity, we find that there are in general three types of stationary
solutions, which we classify below.

Each may be stable or unstable, as we see in the next subsection:
\begin{enumerate}
\item Vacuum solutions. These have $\beta^{(1)}=0$, which is always a stationary
solution.
\item Positive branch solutions. Taking the upper sign in the intensity
equation (\ref{eq:intensity}), this has a positive solution either
if $\Re(c)=\left(c+c^{*}\right)/2<0$ and $|\Im(c)|\le1$, or else
if there is a large enough driving field so that $|c|<1\,$. The corresponding
amplitude is:
\begin{equation}
\beta^{(2)}=\pm\sqrt{1-c\Pi(c)+\left(c^{2}-|c|^{2}\right)/2}.
\end{equation}
This is the only possible solution in the mean-field limit, if one
has no detunings. We show in the next section that this corresponds
to a stationary point~(\ref{minimal_point}) of the quantum P-representation
distribution.
\item Negative branch solutions. Next, investigating the lower sign case
of \ref{eq:intensity}, this has a positive solution if both $\Re(c)<0$,
and there is a small enough driving field so that $|c|>1>|\Im(c)|\,$,
with a resulting amplitude of:
\begin{equation}
\beta^{(3)}=\pm\sqrt{1+c\Pi(c)+\left(c^{2}-|c|^{2}\right)/2}.
\end{equation}
Since we are considering the case of $\Delta_{2}=0$ , we have the
equation:
\begin{equation}
\Re(c)=\frac{\gamma^{(2)}\gamma_{1}^{(1)}+\chi\Delta_{1}}{\left|g\mathcal{E}\right|}.
\end{equation}
Thus $\Re(c)<0$ is possible if we take $\chi\Delta_{1}<-\gamma^{(2)}\gamma_{1}^{(1)}$.
This can occur with an anharmonic term at large detunings, even if
$\gamma_{1}^{(1)}\gg\chi$ as required for the mean-field limit. As
we see below, this occurs as an unstable branch in degenerate parametric
oscillation. This is analogous to the bistable intensity found in
an anharmonic cavity \cite{drummond1980quantum} driven at the fundamental.
In this case the cavity is driven at its second harmonic.
\end{enumerate}

\subsection{Mean-field stability properties}

We are interested in the stability of these steady states. To obtain
this, we calculate the linearized equations for small perturbations,
which reads

\begin{equation}
\frac{d}{d\tau}\left[\begin{array}{c}
\delta\beta\\
\delta\beta^{*}
\end{array}\right]=\left[\begin{array}{cc}
e^{i\theta}\left[-2\left|\beta\right|^{2}-c\right] & e^{i\theta}\left[\left(1-\beta^{2}\right)\right]\\
e^{-i\theta}\left[\left(1-\beta^{*2}\right)\right] & e^{-i\theta}\left[-2\left|\beta\right|^{2}-c^{*}\right]
\end{array}\right]\left[\begin{array}{c}
\delta\beta\\
\delta\beta^{*}
\end{array}\right].
\end{equation}
The corresponding eigenvalues $\lambda_{\pm}$ are
\begin{eqnarray}
\lambda_{\pm} & = & -\Re\left[e^{i\theta}\left[2\left|\beta\right|^{2}+c\right]\right]\pm\\
 & \, & \,\sqrt{\left(\Re\left[e^{i\theta}\left[2\left|\beta\right|^{2}+c\right]\right]\right)^{2}-\left|2\left|\beta\right|^{2}+c\right|^{2}+\left|1-\beta^{2}\right|^{2}}.\nonumber 
\end{eqnarray}

We obtain stability if all eigenvalues are negative, so that there
are two conditions for stability:
\begin{equation}
(a)\,\Re\left[e^{i\theta}\left[2\left|\beta\right|^{2}+c\right]\right]>0\,,
\end{equation}
and:
\begin{equation}
(b)\,\left|2\left|\beta\right|^{2}+c\right|^{2}-\left|1-\beta^{2}\right|^{2}>0.
\end{equation}

We now analyze the different types of solution:

\paragraph*{Type 1 solutions}

For $\beta^{(1)}=0$ the stability condition (a) implies that: 
\begin{equation}
\Re\left[e^{i\theta}c\right]=\frac{\gamma_{1}^{(1)}}{\left|g\right|}>0\,,
\end{equation}
which is always true, and also from condition (b) that the driving
field is below threshold, i.e,

\begin{equation}
\left|c\right|>1.
\end{equation}

Thus, the vacuum solution is stable below threshold and unstable above
threshold.

\paragraph*{Type 2 solutions}

The $\beta^{(2)}$ solutions can occur for either sign of $\Re\left(c\right)$.
\begin{itemize}
\item Firstly, consider the case of $\Re\left(c\right)>0$. For the above
threshold case, condition (a) implies that 
\begin{equation}
\Re\left[e^{i\theta}\left(2\Pi(c)-c^{*}\right)\right]>0,
\end{equation}
which is always true in the region of $\left|c\right|<1$ where the
solution is valid. Condition (b) takes the form of
\begin{equation}
\left[\Pi(c)-\Re(c)\right]\Pi(c)>0,
\end{equation}
which is equivalent to $\left|c\right|<1$. Thus the solution is stable
when $\Re\left(c\right)>0$, $\left|c\right|<1$.
\item Next, consider the case of $\Re\left(c\right)<0$. We find that both
of the stability conditions are always true, provided the solution
exists, which is for $|\Im(c)|\le1$ .
\end{itemize}

\paragraph*{Type 3 solutions}

The $\beta^{(3)}$ solutions only occur for $\Re\left(c\right)<0$.
Condition (a) implies that, for stability,
\begin{equation}
\Re\left[e^{i\theta}\left[-c^{*}-2\Pi(c)\right]\right]>0.
\end{equation}
Condition (b) implies that
\begin{equation}
\left[\Pi(c)+\Re(c)\right]\Pi(c)>0,
\end{equation}
which is only satisfied if $|c|<1$. Since the solution is valid only
if $|c|>1$, the type 3 solution is never stable.

\subsection{Phase-diagram and discrete time symmetry breaking}

\begin{figure}
\centering \includegraphics[width=0.48\textwidth]{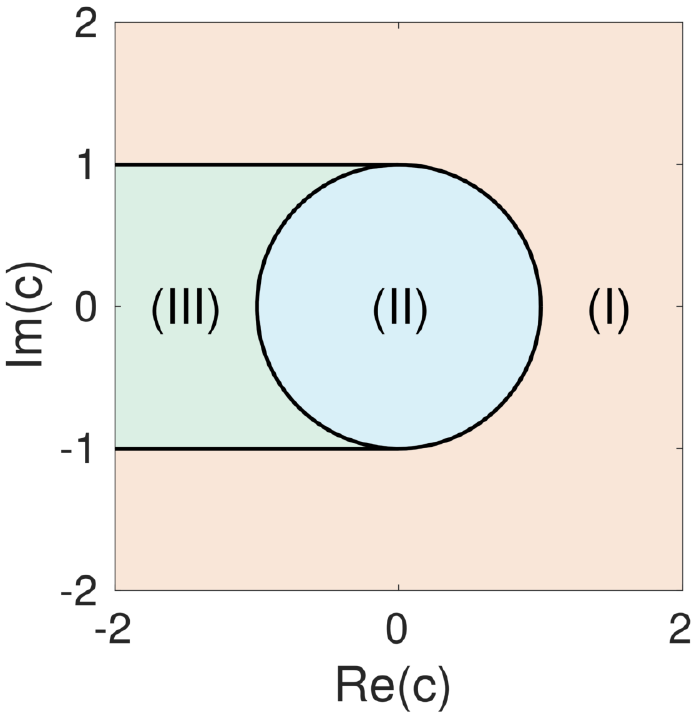}\caption{The phase diagram for mean-field stability. The unique stable vacuum
solution can be observed in region (I). In region (II) above threshold,
there is bistability. Region (III) is where one can observe tristability.}
\label{fig:phase-diagram}
\end{figure}

We have found three types of stationary solutions. Their behavior
changes depending on the complex parameter $c$. This gives a definite
phase-diagram, as shown in figure~\ref{fig:phase-diagram}, with
one phase for $|c|<1$, and two phases for $|c|>1$.
\begin{description}
\item [{(I)}] In the region of $\Re\left(c\right)>0$ we find only one
stationary point, the $\beta^{(1)}$ stable vacuum solution, with
$|c|>1$. This phase also extends to $\Re\left(c\right)<0$, in which
case there is an additional constraint. We find a unique stable vacuum
solution only if $1<|\Im(c)|<|c|$.
\item [{(II)}] There are three stationary points when $|c|<1$, for either
sign of $\Re\left(c\right)$. In this phase the $\beta^{(1)}$ vacuum
solution exists but is unstable, while there are a pair of $\beta^{(2)}$
stable above threshold solutions. This is an example of bistability.
\item [{(III)}] For $\Re\left(c\right)<0$, and $|\Im(c)|<1<|c|,$ there
are five stationary points with $\beta^{(1)}$ stable, $\beta^{(2)}$
stable, and $\beta^{(3)}$ unstable. This is an example of tristability.
\end{description}
In summary, the vacuum steady state $\beta=0$ is stable only at a
driving field below threshold, of $\mathcal{E}<\left|\gamma\right|$,
or $\kappa\mathcal{E}_{2}<\gamma_{2}\left|\gamma\right|$, and the
bistable states occur at large driving field, when $\kappa\mathcal{E}_{2}>\gamma_{2}\left|\gamma\right|$.
There is also a tristable regime below threshold in which the vacuum
state is stable, but there are stable solutions of finite amplitude
as well. Otherwise, the real part of the eigenvalue will be positive,
which means that the solutions become unstable in the corresponding
parameter region.

Hence, in general there can be up to five stationary solutions. The
nonvanishing solutions correspond to an output electromagnetic field
of 
\begin{equation}
E\propto\pm\sin\left(\omega\left(t-t_{1}\right)\right)\,,
\end{equation}
where $t_{1}$ is a time-origin that depends on the details of the
input and output coupling.

The stable $\beta^{(2)}$ solutions exist in either bistable (phase
II) or tristable (phase III) regimes. These solutions always come
in pairs, whose two solutions differ by a time translation, since:
\begin{eqnarray}
-\sin\left(\omega\left(t-t_{1}\right)\right) & = & \sin\left(\omega\left(t-t_{1}\right)-\pi\right)\nonumber \\
 & = & \sin\left(\omega\left(t-t_{2}\right)\right).
\end{eqnarray}
Here $t_{2}=t_{1}+\pi/\omega$, which defines a second discrete time
origin. Thus, the system can relax to a steady-state that corresponds
to either time origin. This is called spontaneous discrete time symmetry
breaking. However, this behavior is only possible if these are stable
solutions.

These results imply that in the mean-field limit, the above-threshold
steady-state solutions have a spontaneous discrete time-translation
symmetry-breaking. However, so far we have ignored quantum fluctuations,
which we turn to next.

\section{Quantum steady-state and tunneling\label{sec:Quantum-steady-state}}

In the mean field picture, once the subharmonic mode reaches a stable
state, no further dynamics will be observed. This picture is not accurate
once fluctuations are taken into account. In the presence of either
classical (thermal noise) \cite{woo1971ieee_thermaltunnel,Bryant_JAP1987_thermaltunnel}
or quantum fluctuations \cite{reid1992effect}, a steady state can
reach the other steady state of the multistable solutions predicted
by mean field theory. Beyond mean field theory, where damping and
noise is included, the degenerate parametric oscillator obeys a master
equation, where the steady state solutions and the switching rate
between these states are obtained either by solving the corresponding
Fokker-Planck equation \cite{Vogel_PRA1988_quantumtunnel,drummond1989quantum,kinsler1991quantum}
or solving for the zero eigenvalue and its eigenvector of the super-operator
dictating the time evolution in a master equation, expanded over a
basis representation such as the Fock state basis\cite{Risken_PRA1987_quantumtunnel,Risken_PRA1988_quantumtunnel,kinsler1991quantum,Casteels_PRA2016,Rodriguez_PRL2017}.

Hence, we now turn to the full quantum behavior of this system. This
allows us to derive the full steady-state quantum statistics, and
to demonstrate that spontaneous symmetry breaking has an exponentially
long lifetime at large photon number. We note that, unlike the mean-field
solutions, these full quantum solutions require the inclusion of off-diagonal
coherent projectors, which increase the phase-space dimensionality
compared to the classical, mean-field behaviour.

\subsection{Steady-state quantum solutions}

In the zero temperature case, the steady-state solution of the scaled
Fokker-Planck equation~(\ref{FPE_scaled}) can be readily found by
using the method of potential equations~\cite{graham1971fluctuations,graham1971generalized,risken1972solutions,seybold1974theory}
\begin{equation}
P_{1}\left(\vec{\beta}\right)=N\exp\left[-\Phi\left(\vec{\beta}\right)\right],\label{steady-state}
\end{equation}
where $N$ is a normalization constant and $\Phi$ satisfies 
\begin{eqnarray}
\frac{\left(1-\beta^{2}\right)}{2n}\frac{\partial\Phi}{\partial\beta}=(c-\frac{1}{n})\beta-\left(1-\beta^{2}\right)\beta^{+}-d,\nonumber \\
\frac{\left(1-\beta^{+2}\right)}{2n}\frac{\partial\Phi}{\partial\beta^{+}}=(c^{*}-\frac{1}{n})\beta^{+}-\left(1-\beta^{+2}\right)\beta-d^{*}.\label{eq:for-potential}
\end{eqnarray}
These differential equations can be obtained by inserting the form~(\ref{steady-state})
into the Fokker-Planck equation~(\ref{FPE_scaled}) and setting $\partial P_{1}/\partial\tau=0$.

It is simplest to proceed by introducing a shifted $c$ parameter:
\begin{equation}
\tilde{c}=c-\frac{1}{n},
\end{equation}
where $\tilde{c}\rightarrow c$ in the classical limit of $n\rightarrow\infty$.
Including quantum noise, the exact steady-state solution is defined
by the potential: 
\begin{eqnarray}
\Phi\left(\vec{\beta}\right) & = & -n\left[\beta^{+}\beta+\tilde{c}\ln(1-\beta^{2})+d\ln\left(\frac{1+\beta}{1-\beta}\right)+h.c.\right],\label{potential_beta_full}
\end{eqnarray}
so that in the scaled coherent space: 
\begin{eqnarray}
P_{S}\left(\vec{\beta}\right) & = & N\left[(1+\beta)^{\tilde{c}+d}(1-\beta)^{\tilde{c}-d}\right.\nonumber \\
 &  & \times\left.(1+\beta^{+})^{\tilde{c}^{\ast}+d^{*}}(1-\beta^{+})^{\tilde{c}^{\ast}-d^{*}}\exp(2\beta^{+}\beta)\right]^{n}.
\end{eqnarray}

This gives the exact zero-temperature solution for the steady-state
of the density matrix, provided it is accompanied by a choice of contours
that leads to a solution that is bounded, which we treat in detail
in the next subsection. While formally similar to previous solutions,
we note that all the parameters here can have complex values, which
is necessary when treating the physics of recent quantum circuit experiments.

In the tunneling calculation below, we assume that the resonant driving
field is only added on mode $a_{2}$ so that $\mathcal{E}_{1}=0$
and $\Delta_{2}=0$, which is the situation in recent experiments~\cite{leghtas2015confining}.
In this case, the exact steady-state potential is: 
\begin{equation}
\Phi\left(\vec{\beta}\right)=-n\left[\beta^{+}\beta+\tilde{c}\ln(1-\beta^{2})+h.c.\right],\label{potential_beta}
\end{equation}
and the probability distribution is:
\begin{equation}
P_{S}\left(\vec{\beta}\right)=N\left[(1-\beta^{2})^{\tilde{c}}(1-\beta^{+2})^{\tilde{c}^{\ast}}\exp(2\beta^{+}\beta)\right]^{n}.
\end{equation}

The solution above is a time-symmetric mixed state, which includes
an equal probability of observing either of the two possible output
amplitudes. We now wish to calculate the rate at which the system
can switch from one phase to the other, i.e., the rate at which time
symmetry is restored once it is broken by an observation of one or
the other of the two possible output amplitudes.

For simplicity in this discussion of tunneling, we will assume $\Re(\tilde{c})>0$
and in the phase (II) bistable region. We get potential solutions
vanishing on boundaries when $\Re(\tilde{c})>0$, as shown in this
work. In the case that $\Re(\tilde{c})<0$, the potential diverges
on the boundaries. This means that the physics of tunneling may be
very different, and requires a different type of manifold for its
treatment, which we analyze elsewhere.

To this end, we discuss the physical interpretation of the parameter
region $\Re(\tilde{c})>0$. We note that $\tilde{c}=\left(\gamma-g\right)/\left(gn\right)$,
$\gamma=\gamma_{1}^{(1)}+i\Delta_{1}$ and $g=\gamma^{(2)}+i\chi$.
As a result, we have
\begin{equation}
\tilde{c}=\frac{\gamma^{(2)}(\gamma_{1}^{(1)}-\gamma^{(2)})-\chi(\chi-\Delta_{1})}{\mathcal{E}\sqrt{(\gamma^{(2)})^{2}+\chi^{2}}}-i\frac{\gamma_{1}^{(1)}\chi-\gamma^{(2)}\Delta_{1}}{\mathcal{E}\sqrt{(\gamma^{(2)})^{2}+\chi^{2}}}.\label{eq:c}
\end{equation}
 Thus we will have $\gamma^{(2)}(\gamma_{1}^{(1)}-\gamma^{(2)})-\chi(\chi-\Delta_{1})>0$
in the region of $\Re(\tilde{c})>0$, which means that typically one
has $\gamma_{1}^{(1)}>\gamma^{(2)}$, although there is also a nonlinear
coupling $\chi$ and a detuning $\Delta_{1}$ which can change this
relationship. We see that in the case that the detuning $\Delta_{1}$
is negligible, the linear damping rate is larger than the nonlinear
coupling when $\Re\left(\tilde{c}\right)>0$.

\subsection{Tunneling regime with $\Re(\tilde{c})>0$\label{sec:Manifold-with-positive-c}}

Geometrically, we can regard the quantum dynamics as occurring via
a distribution function defined on a two-dimensional manifold embedded
in a four-dimensional complex space. In this paper, we focus on the
tunneling-dominated regime of $\Re(\tilde{c})>0$, i.e. the region
of large single-photon loss and small nonlinear coupling, so that
the potential function vanishes at the boundaries of a probability
domain defined by square boundaries at $\beta=\pm1$ or $\beta^{+}=\pm1$.
The manifold must also includes the vacuum state at $\beta=\beta^{+}=0$,
which is the starting point of any dynamical experiment.

We expect tunneling between minima of the potential as in earlier
work~\cite{drummond1989quantum,kinsler1991quantum}. However, in
this earlier work, the parameters were real and there was no anharmonicity.
In the present case, the probability domain that includes these minima
is no longer necessarily a plane with real values of $\beta,\beta^{+}$.
We now analyze the locations of these minima in the four dimensional
space of coherent amplitudes.

To find the stable points, we will solve two equations analogous to
the mean-field stationarity conditions, but generalized to four dimensions:
\begin{eqnarray}
\Phi_{1} & \equiv & \frac{\partial\Phi}{\partial\beta}=n\left[-2\beta^{+}+\frac{2\tilde{c}\beta}{1-\beta^{2}}\right]=0,\nonumber \\
\Phi_{2} & \equiv & \frac{\partial\Phi}{\partial\beta^{+}}=n\left[-2\beta+\frac{2\tilde{c}^{\ast}\beta^{+}}{1-\beta^{+2}}\right]=0.\label{to_stable_point}
\end{eqnarray}
 If $\beta$ and $\beta^{+}$ are nonzero real numbers, we see that
$-2\beta^{+}$ is real, but in general $2\tilde{c}\beta/(1-\beta^{2})$
is complex, so the equations~(\ref{to_stable_point}) can't be satisfied.
This means that there is generally at least one complex number in
$\beta$ and $\beta^{+}$ for nontrivial solutions of the stationary
points of the potential.

The potential function~(\ref{potential_beta}) could be complex because
$\tilde{c}$, $\beta$ and $\beta^{+}$ are complex numbers. The stationary
points obtained by equations~(\ref{to_stable_point}) are divided
into three types: the origin solution ($\beta=\beta^{+}=0$), the
classical solutions ($\beta^{+}=\beta^{*}$) and the nonclassical
solutions ($\beta^{+}=-\beta^{*}$). For all three types of solution,
the potential functions~(\ref{potential_beta}) are always real.
This means that we can study the stationary points in their neighborhoods
to find whether they are local minima, maxima, or saddle points. In
each case, we assume that required relations define a locally planar
surface in a neighborhood of the solution, in order to define the
derivatives.

\subsection{Local stationary points\label{subsec:Local-stationary-points}}

As mentioned above, we find three types of solutions on solving equations~(\ref{to_stable_point}).
It is common to use the Hessian matrix to determine whether the roots
are local minima, maxima, or saddle points~\cite{binmore2002calculus}.
The Hessian matrix is defined by the second derivatives of the potential
function:
\begin{equation}
M=\left[\begin{array}{cc}
\Phi_{11} & \Phi_{12}\\
\Phi_{21} & \Phi_{22}
\end{array}\right],
\end{equation}
where 
\begin{eqnarray}
\Phi_{11} & \equiv & \frac{\partial^{2}\Phi}{\partial\beta^{2}}=\frac{2\tilde{c}n(1+\beta^{2})}{(1-\beta^{2})^{2}},\quad\Phi_{22}\equiv\frac{\partial^{2}\Phi}{\partial\beta^{+2}}=\frac{2\tilde{c}^{*}n(1+\beta^{+2})}{(1-\beta^{+2})^{2}},\nonumber \\
\Phi_{12} & \equiv & \frac{\partial^{2}\Phi}{\partial\beta^{+}\partial\beta}=-2n=\Phi_{21}.
\end{eqnarray}
If the Hessian matrix is positive definite at a stationary point $\vec{\beta}$,
$\vec{\beta}$ is an isolated local minimum of the potential function
$\Phi(\vec{\beta})$. For a $2\times2$ matrix, positive definite
is equivalent to a positive determinant $\left|M\right|>0$ and a
positive trace $\Tr(M)>0$. Similarly, the Hessian matrix is negative
definite at a stationary point when it is an isolated local maximum,
which is equivalent to a positive determinant $\left|M\right|>0$
and a negative trace $\Tr(M)<0$. At a saddle point, the Hessian matrix
has both positive and negative eigenvalues, which leads to a negative
determinant $\left|M\right|<0$~\cite{binmore2002calculus,stewart2009multivariable}.

The first type of solution is at the origin, $\beta=\beta^{+}=0$.
The potential is simply $\Phi^{(o)}\equiv\Phi(0,0)=0$, and in this
case, the second derivatives are 
\begin{eqnarray}
\Phi_{11}^{(o)} & = & \left.\frac{2\tilde{c}(1+\beta^{2})}{(1-\beta^{2})^{2}}\right|{}_{\beta=0}=2\tilde{c}n,\nonumber \\
\Phi_{22}^{(o)} & = & \left.\frac{2\tilde{c}(1+\beta^{+2})}{(1-\beta^{+2})^{2}}\right|{}_{\beta^{+}=0}=2\tilde{c}^{*}n,\nonumber \\
\Phi_{12}^{(o)} & = & -2n.\label{eq:sec-der-type1}
\end{eqnarray}
Therefore, the Hessian matrix determinant is obtained as 
\begin{equation}
\left|M^{(o)}\right|=\left|\begin{array}{cc}
\Phi_{11}^{(o)} & \Phi_{12}^{(o)}\\
\Phi_{21}^{(o)} & \Phi_{22}^{(o)}
\end{array}\right|=4n^{2}(|\tilde{c}|^{2}-1).\label{eq:det-type1}
\end{equation}
If $|\tilde{c}|<1$, we have $\left|M^{(o)}\right|<0$, which means
that the origin point is a saddle point. This is generally stable
below threshold, and unstable above threshold, as expected from the
previous mean-field analysis.

As we will see in the following, other minima as well as the quantum
tunneling only occur when $|\tilde{c}|<1$, which means that this
first type of solution plays the role of a saddle point in understanding
quantum tunneling. In subharmonic generators, the bistable solutions
only take place above threshold. In our calculations, bistable solutions
are obtained as double minima in the manifold, occurring in the parameter
region of $|\tilde{c}|<1$. It is also possible to have tristability,
as explained in the mean-field section.

\subsection{Classical stable points}

The second type of stable point is $\beta^{+}=\beta^{\ast}=r\exp(-i\varphi)\neq0$.
These conditions would correspond to a coherent state projector, so
we term them classical stable points, and they closely match the corresponding
stable mean-field solutions for $n\gg1$. For simplicity, we assume
that we have a bistable or phase II situation, rather than the more
complex tristable phase III situation.

In this case equations~(\ref{to_stable_point}) can be transformed
into 
\begin{equation}
re^{-i\varphi}=\frac{\tilde{c}re^{i\varphi}}{1-r^{2}e^{2i\varphi}},
\end{equation}
and therefore, 
\begin{equation}
r^{2}=e^{-2i\varphi}-\tilde{c}.
\end{equation}
Because $r$ is real, we have 
\begin{eqnarray}
\sin(2\varphi)=-\Im(\tilde{c}),\quad r^{2}=\pm\sqrt{1-\Im(\tilde{c})^{2}}-\Re(\tilde{c})>0.
\end{eqnarray}
Taking the positive sign to obtain the bistable region, the condition
$r^{2}>0$ is equivalent to $|\tilde{c}|<1$. This means that $r^{2}=\left|\beta\right|^{2}<1$
as well. We note that for small values of $n$, the exact phase boundaries
are modified due to the fact that $c\neq\tilde{c}.$ In fact it is
better to regard this as a somewhat fuzzy criterion at small $n$,
since quantum fluctuations tend to broaden these phase distinctions.

Labelling these stationary points as $\beta^{(c)}$, $\beta^{(c)+}$,
we finally get 
\begin{eqnarray}
\beta^{(c)} & = & \pm\left[1+\left(\tilde{c}^{2}-|\tilde{c}|^{2}\right)/2-\tilde{c}\Pi(\tilde{c})\right]^{1/2},\nonumber \\
\beta^{(c)+} & = & \pm\left[1+\left(\tilde{c}^{*2}-|\tilde{c}|^{2}\right)/2-\tilde{c}^{*}\Pi(\tilde{c})\right]^{1/2}.\label{minimal_point}
\end{eqnarray}
Here we have set $\Pi(\tilde{c})=\sqrt{1-\Im\left(\tilde{c}\right)^{2}}$
as in the mean-field analysis, while $\beta^{(c)}$ and $\beta^{(c)+}$
are generally complex. These solutions correspond to the mean-field
solutions above threshold obtained previously, except that now we
use the quantum noise modified value of the coupling $\tilde{c},$
rather than its mean-field value. If $\tilde{c}$ is a real number,
the results will reduce to $(\beta^{(c)},\beta^{(c)+})=(\pm\sqrt{1-\tilde{c}},\pm\sqrt{1-\tilde{c}})$,
which corresponds to the first line of equation~(4.7) in Ref.~\cite{kinsler1991quantum}
with $n=\mu/g^{2}$, $\tilde{c}=\sigma/\mu$, $\alpha_{0}=\sqrt{n}\beta$,
with no anharmonic term.

In this classical stationary point case, the second derivatives are
\begin{eqnarray}
\Phi_{11}^{(c)} & = & \frac{n}{\tilde{c}}\frac{4+\tilde{c}^{2}-|\tilde{c}|^{2}-2\tilde{c}\Pi(\tilde{c})}{\left[\Im\left(\tilde{c}\right)-\Pi(\tilde{c})\right]^{2}},\nonumber \\
\Phi_{22}^{(c)} & = & \frac{n}{\tilde{c}^{\ast}}\frac{4+\tilde{c}^{\ast2}-|\tilde{c}|^{2}-2\tilde{c}^{\ast}\Pi(\tilde{c})}{\left[\Im\left(\tilde{c}\right)+\Pi(\tilde{c})\right]^{2}},\nonumber \\
\Phi_{12}^{(c)} & = & -2n.\label{eq:sec-der-type2}
\end{eqnarray}
Hence, the determinant of Hessian matrix is obtained, 
\begin{equation}
\left|M^{(c)}\right|=\frac{16n^{2}}{|\tilde{c}|^{2}}\Pi(\tilde{c})[\Pi(\tilde{c})-\Re(\tilde{c})].\label{eq:det-type2}
\end{equation}
We expect that the classical stable points are only valid with the
condition $|\tilde{c}|<1$. It is directly checked that $\left|M^{(c)}\right|>0$
and the real part $Re[\Phi_{11}^{(c)}]>0$. We note that $\Phi_{11}^{(c)}=\Phi_{22}^{(c)\ast}$,
which leads to $\Tr[M^{(c)}]>0$. Thus, these stable points are minima
in the quantum potential, that is, they are local attractors. As the
bistable states only occur above threshold, the condition $|\tilde{c}|<1$
corresponds to the threshold value in degenerate parametric oscillators.

\subsection{Complex phase-space manifold\label{subsec:Complex-phase-space-manifold}}

As introduced at the beginning of the section~\ref{sec:Manifold-with-positive-c}
and in earlier work~\cite{drummond1989quantum,kinsler1991quantum},
in order to study quantum tunneling analytically, we need to find
a two-dimensional manifold embedded in a four-dimensional complex
space. The potential function defined on this manifold vanishes at
the boundaries $\beta=\pm1$ or $\beta^{+}=\pm1$. The manifold must
also include the vacuum state at $\beta=\beta^{+}=0$, as well as
the classical stable points~(\ref{minimal_point}), so that quantum
tunneling can take place. In order to define them as local saddle
points or minima, we assume that on the manifold there is a locally
planar surface in a neighborhood of the solutions.

\begin{figure}
\centering \includegraphics[width=0.48\textwidth]{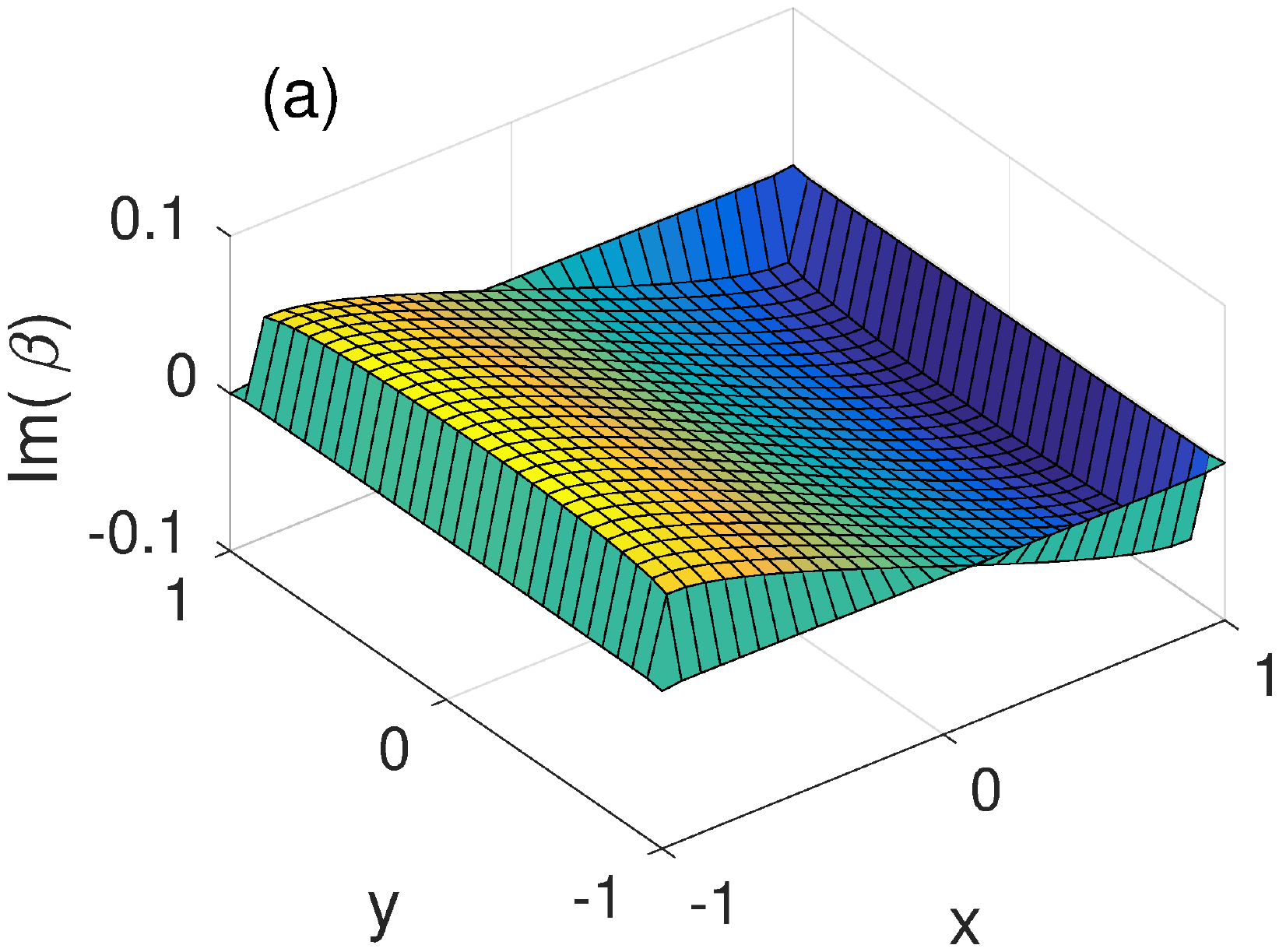} \includegraphics[width=0.48\textwidth]{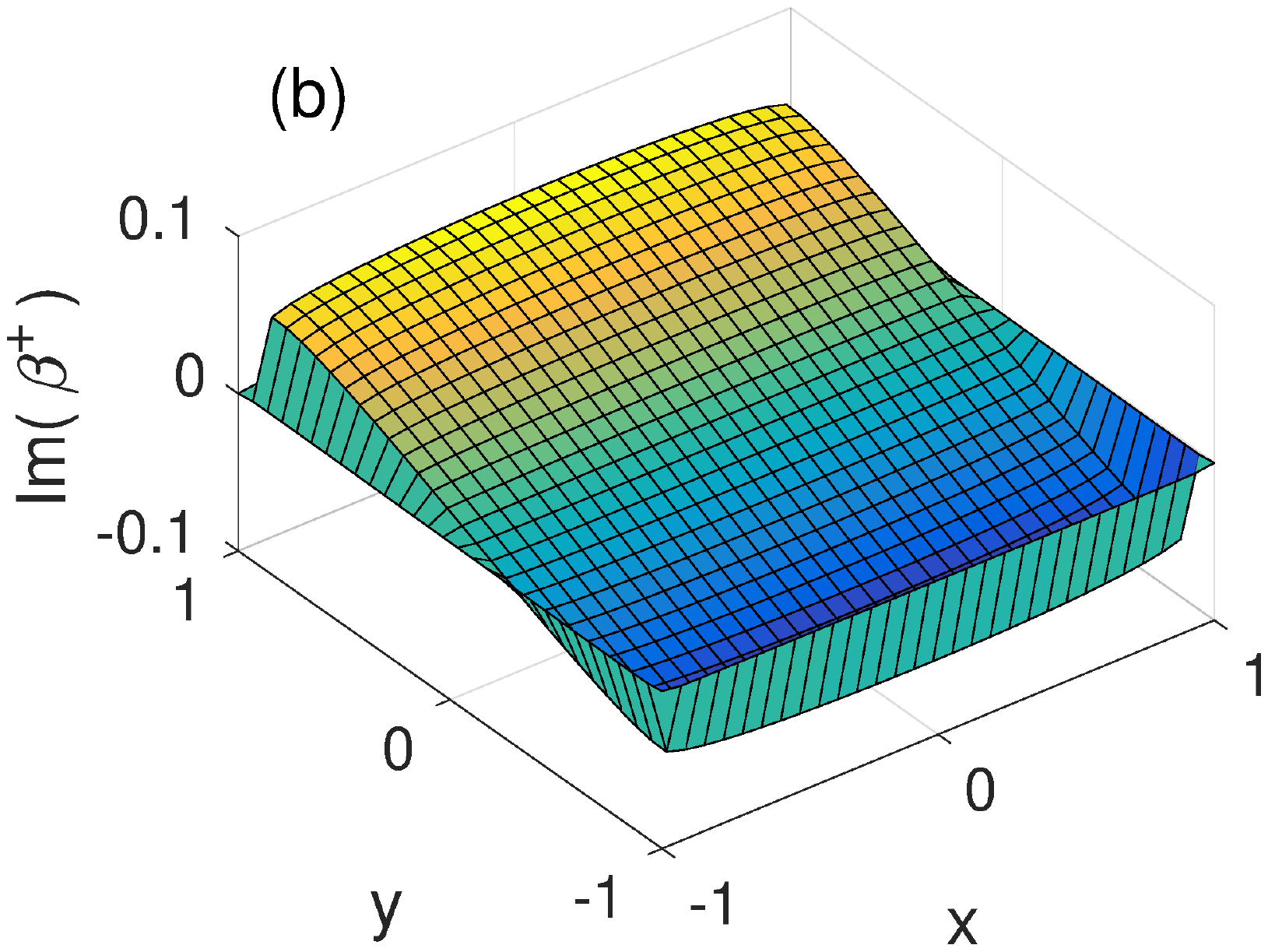}
\caption{The manifold where we define our Fokker-Planck equation and the potential.
The manifold has been parameterized~(\ref{eq:manifold}). Because
it is embedded in a 4D space, we can only plot the manifold with one
dimension omitted. Figure (a) shows the manifold with real variables
$(x,y,Im(\beta))$ where the imaginary part of $\beta^{+}$ is omitted,
and figure (b) shows $(x,y,Im(\beta^{+}))$ where the imaginary part
of $\beta$ is omitted. In these figures, $\tilde{c}=0.33+0.17i$,
$n=3$ and $p=0.01$.}
\label{fig:manifold}
\end{figure}

Given these considerations, we define a curved surface through the
stable points, $\beta^{(c)}=e^{i\varphi}\left|\beta^{(c)}\right|$,
parameterized as: 
\begin{eqnarray}
\beta & =x\left(1+i\tan\left(\varphi\right)\right),\quad\beta^{+} & =y\left(1-i\tan\left(\varphi\right)\right).
\end{eqnarray}

For large $x,y$ we want to include the real boundaries such that
$\beta=\pm1$, $\beta^{+}=\pm1$, are on the manifold. Therefore,
we can modify this as: 
\begin{eqnarray}
\beta & =x+ix\tan\left(\varphi\right)\cos^{p}\left(x\pi/2\right)\cos^{p}\left(y\pi/2\right),\nonumber \\
\beta^{+} & =y-iy\tan\left(\varphi\right)\cos^{p}\left(x\pi/2\right)\cos^{p}\left(y\pi/2\right).\label{eq:manifold}
\end{eqnarray}
In the limit of $p\rightarrow0$ this gives the correct behavior of
the required manifold, as a tilted plane which is cutoff at the edges
to give the square manifold with vanishing boundaries.

This manifold is plotted in the figure~\ref{fig:manifold}, for $p=0.01$.
One could also modify the cutoff function to give different behavior
in the classical and quantum directions, but we only wish to consider
the simplest case here. 
\begin{figure}
\centering \includegraphics[width=0.48\textwidth]{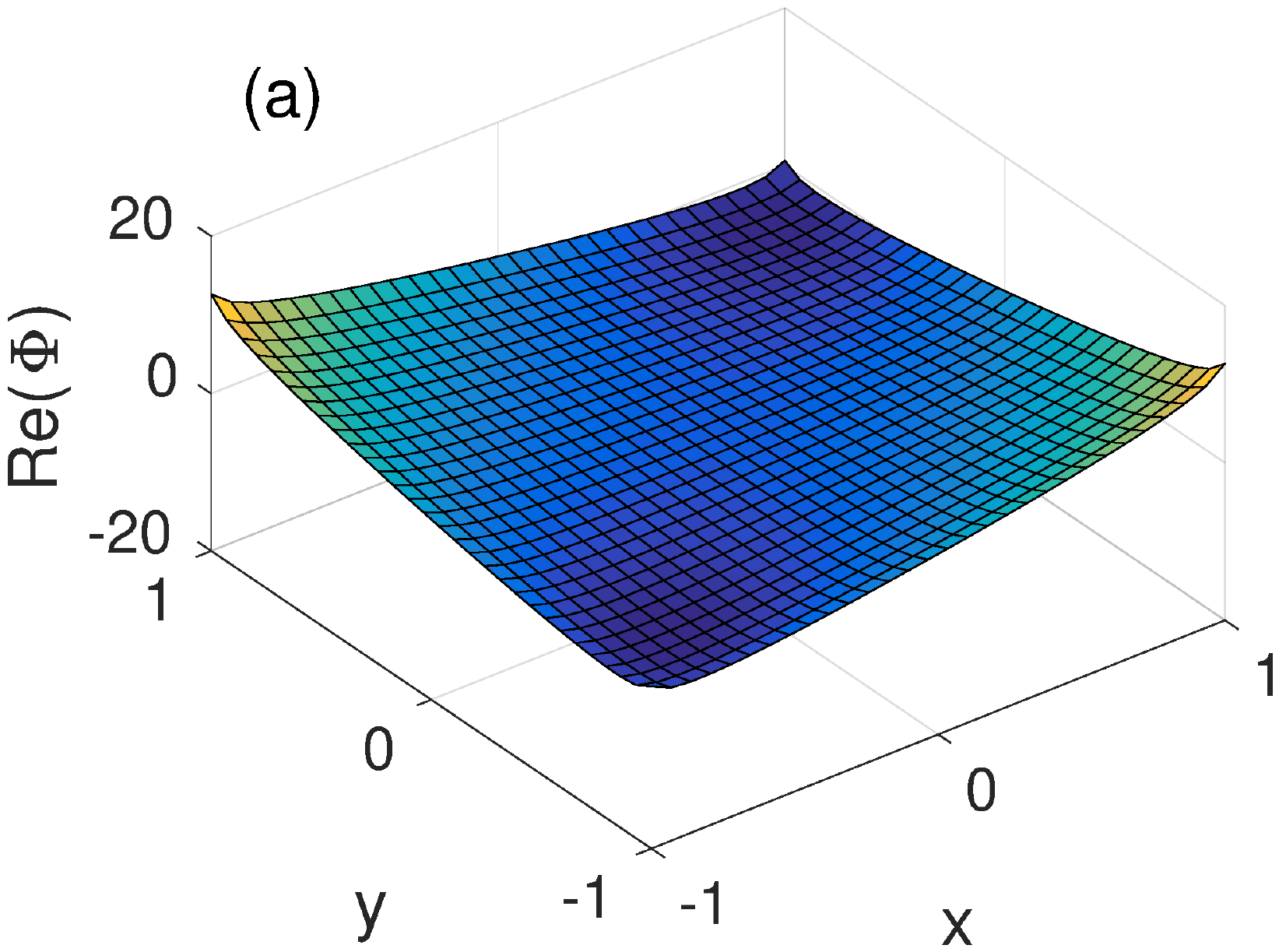} \includegraphics[width=0.48\textwidth]{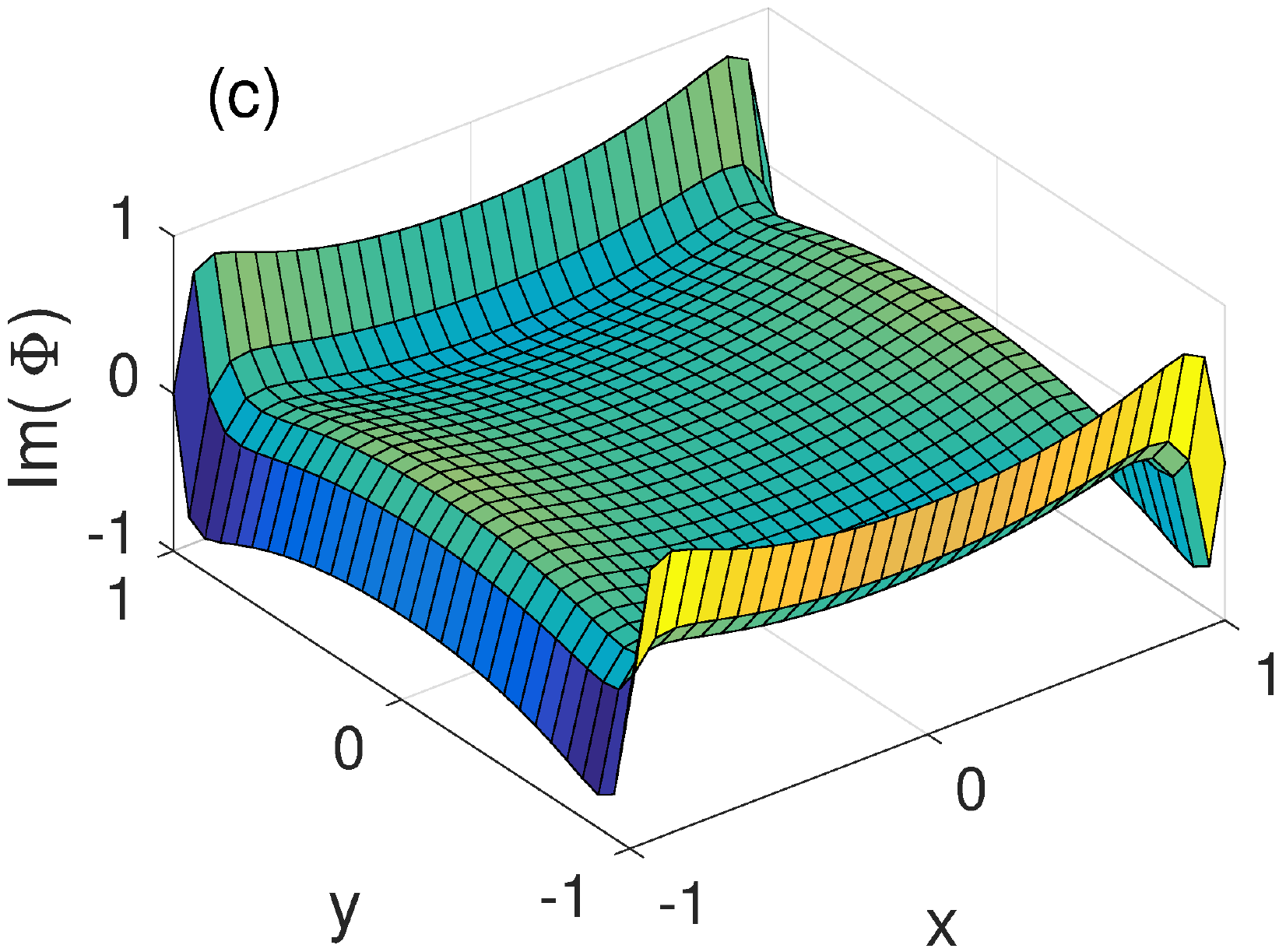}\\
 \hspace*{\fill} \includegraphics[width=0.48\textwidth]{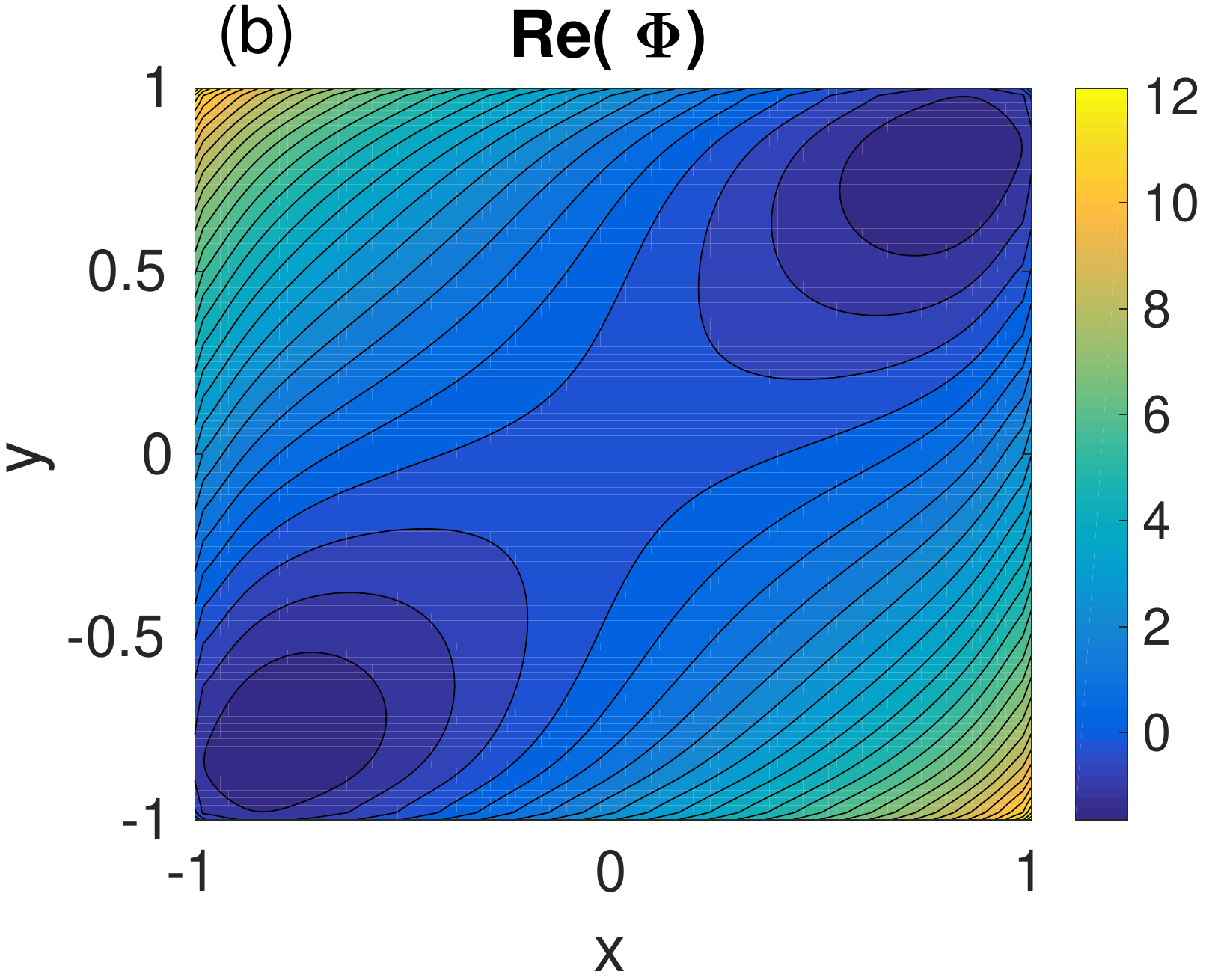}
\includegraphics[width=0.48\textwidth]{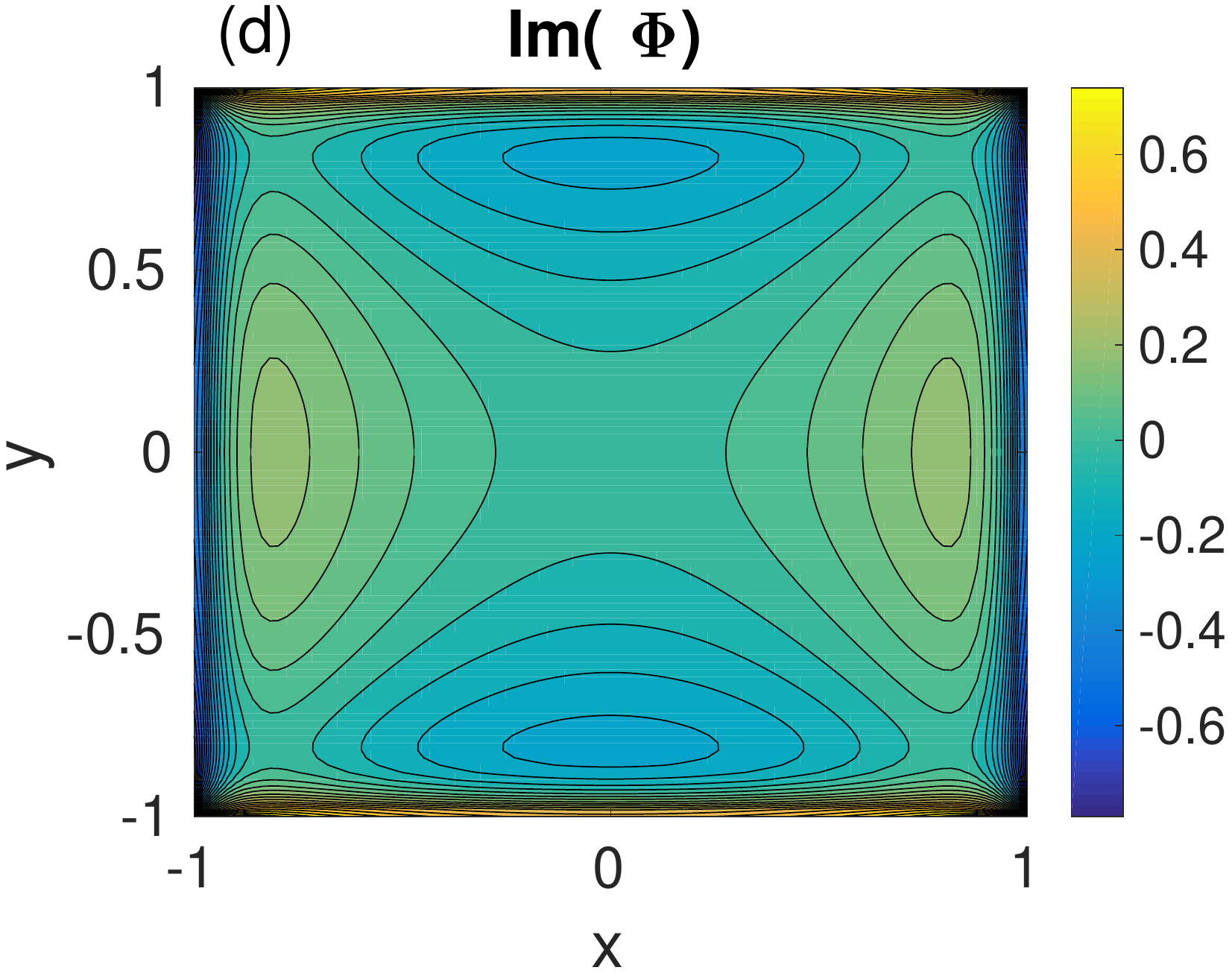} \caption{The potential $\Phi(\vec{\beta})$~(\ref{potential_beta}) on the
manifold~(\ref{eq:manifold}). Figure (a) shows the real part of
$\Phi(\vec{\beta})$ with the parameterized variables $(x,y)$, and
figure (b) shows the related contour figure of $Re[\Phi(\vec{\beta})]$.
Figure (c) shows the image part of $\Phi(\vec{\beta})$ with $(x,y)$,
and the figure (d) shows the related contour figure of $Im[\Phi(\vec{\beta})]$.
In these figures, $\tilde{c}=0.33+0.17i$, $n=3$ and $p=0.01$.}
\label{fig:potential} 
\end{figure}

On this manifold~(\ref{eq:manifold}), we show the potential $\Phi(\vec{\beta})$~(\ref{potential_beta})
in figure~\ref{fig:potential}. We can obtain the local minima and
the saddle point from the real part of the potential in figure~\ref{fig:potential}~(a)
and (b). In the neighborhood of these points, figures~\ref{fig:potential}~(c)
and (d) show that $Im(\Phi)=0$. Thus it is valid to define local
minima and a saddle point just as with real potentials.

We also show that quantum tunneling will take place between these
classical minima through the saddle point in the~\ref{sec:Tunneling-rate-for-complex},
where we generalized the potential-barrier approximation to complex
cases and derived the tunneling time for a complex Fokker-Planck equation.

\subsection{Quantum stable points}

There is another possible stationary value of the potential, which
is for$\beta=-\beta^{+\ast}$. These would correspond to a superposition
of distinct coherent states, which is a uniquely quantum effect, so
we term them quantum stable points. By labelling these points as $\beta^{(q)}$
and $\beta^{(q)+}$, we finally obtain
\begin{eqnarray}
\beta^{(q)} & = & \pm\left[1+\left(\tilde{c}^{2}-|\tilde{c}|^{2}\right)/2+\tilde{c}\Pi(\tilde{c})\right]^{1/2},\nonumber \\
\beta^{(q)+} & = & \mp\left[1+\left(\tilde{c}^{*2}-|\tilde{c}|^{2}\right)/2+\tilde{c}^{*}\Pi(\tilde{c})\right]^{1/2},\label{eq:root-type3}
\end{eqnarray}
with $|\tilde{c}|<1$, i.e. above threshold. If $c$ is a real number,
the results reduce to $(\beta^{(q)},\beta^{(q)+})=(\pm\sqrt{1+\tilde{c}},\mp\sqrt{1+\tilde{c}})$,
which corresponds to the second line of equation~(4.7) in Ref.~\cite{kinsler1991quantum},
where there is no anharmonic term.

In this case, the second derivatives are 
\begin{eqnarray}
\Phi_{11}^{(q)} & = & \frac{n}{\tilde{c}}\frac{4+\tilde{c}^{2}-|\tilde{c}|^{2}+2\tilde{c}\Pi(\tilde{c})}{\left[\Im\left(\tilde{c}\right)+\Pi(\tilde{c})\right]^{2}},\nonumber \\
\Phi_{22}^{(q)} & = & \frac{n}{\tilde{c}^{\ast}}\frac{4+\tilde{c}^{\ast2}-|\tilde{c}|^{2}+2\tilde{c}^{\ast}\Pi(\tilde{c})}{\left[\Im\left(\tilde{c}\right)-\Pi(\tilde{c})\right]^{2}},\nonumber \\
\Phi_{12}^{(q)} & = & -2n.\label{eq:sec-der-type3}
\end{eqnarray}
Hence, the Hessian determinant is 
\begin{eqnarray}
\left|M^{(q)}\right| & = & \frac{16n^{2}}{|\tilde{c}|^{2}}\Pi(\tilde{c})[\Pi(\tilde{c})+\Re(\tilde{c})].\label{eq:det-type3}
\end{eqnarray}
Considering that $|\tilde{c}|<1$, it is easily checked that $\left|M^{(q)}\right|>0$
and $Re[\Phi_{11}^{(q)}]>0$. Because $\Phi_{22}^{(q)}$ is the complex
conjugate of $\Phi_{11}^{(q)}$, the trace of Hessian matrix is then
positive: $\Tr[M^{(q)}]>0$. Thus, these stable points are also expected
to be minima of the potential.

We note that $|\beta^{(q)}|>1$ since $|\tilde{c}|<1$, so these are
farther from the origin than the classical minima, and in fact outside
the boundaries of the stable manifold considered here. Therefore,
tunneling occurs between the stable classical minima through the saddle
point at the origin. This type of quantum stable point becomes important
for extremely strong coupling, and will be treated in detail elsewhere.

\subsection{Tunneling rate\label{subsec:Tunneling-rate}}

In order to calculate the tunneling rate between the classical minima
through the saddle point at the origin, we use a transformation to
define variables $(u,v)=F(\beta,\beta^{+})$ where the classical minimal
points are placed on the axis of $u$. With these new variables, the
diffusion coefficient is a constant. This simplifies the calculation
of the tunneling time via the potential-barrier approximation~\cite{kramers1940brownian,landauer1961frequency,drummond1989quantum,kinsler1991quantum},
which is generalized for complex cases in~\ref{sec:Tunneling-rate-for-complex}.

Here we consider the transformation introduced in~\ref{sec:Tunneling-rate-for-complex},
and combine equations~(\ref{eq:transform-1}) and (\ref{eq:transform-2})
to give: 
\begin{eqnarray}
u & = & e^{-i\theta/2}e^{i\phi}\sin^{-1}\beta+e^{i\theta/2}e^{-i\phi}\sin^{-1}\beta^{+},\nonumber \\
v & = & e^{-i\theta/2}e^{-i\phi}\sin^{-1}\beta-e^{i\theta/2}e^{i\phi}\sin^{-1}\beta^{+}.\label{eq:transform}
\end{eqnarray}
The inverse transformation of equation~(\ref{eq:transform}) is
\begin{eqnarray}
\beta & = & \sin\left[\Upsilon_{+}\right],\quad\beta^{+}=\sin\left[\Upsilon_{-}\right],
\end{eqnarray}
with the notation that 
\begin{eqnarray}
\Upsilon_{+} & \equiv & \frac{e^{i(\theta/2+\phi)}u+e^{i(\theta/2-\phi)}v}{2\cos2\phi},\nonumber \\
\Upsilon_{-} & \equiv & \frac{e^{-i(\theta/2+\phi)}u-e^{-i(\theta/2-\phi)}v}{2\cos2\phi}.
\end{eqnarray}
As explained in the~\ref{sec:Tunneling-rate-for-complex},
$\phi=\psi-\theta/2$ with $\sin^{-1}\beta^{(c)}=re^{i\psi}$. Here
$e^{i\theta}=g/|g|=n/\epsilon$ has been introduced in the section~\ref{sec:Adiabatic-Elimination}.
In the following, we will express the classical stable points $\beta^{(c)}$
with the newly introduced variables $u$ and $v$. Then we can study
the tunneling rate for our system by applying the analytic tunneling
time result~(\ref{eq:tunneling_complex}) obtained in the~\ref{sec:Tunneling-rate-for-complex}. 

Considering the manifold (\ref{eq:manifold}) where our potential
is defined on, the variables $(u,v)$ can be parameterized directly,
\begin{eqnarray}
u & =e^{-i\theta/2}e^{i\phi}\sin^{-1}\left[x+ix\tan\left(\varphi\right)\cos^{p}\left(x\pi/2\right)\cos^{p}\left(y\pi/2\right)\right]\nonumber \\
 & \quad+e^{i\theta/2}e^{-i\phi}\sin^{-1}\left[y-iy\tan\left(\varphi\right)\cos^{p}\left(x\pi/2\right)\cos^{p}\left(y\pi/2\right)\right],\\
v & =e^{-i\theta/2}e^{-i\phi}\sin^{-1}\left[x+ix\tan\left(\varphi\right)\cos^{p}\left(x\pi/2\right)\cos^{p}\left(y\pi/2\right)\right]\nonumber \\
 & \quad-e^{i\theta/2}e^{i\phi}\sin^{-1}\left[y-iy\tan\left(\varphi\right)\cos^{p}\left(x\pi/2\right)\cos^{p}\left(y\pi/2\right)\right],
\end{eqnarray}
 with $p\to0$. It is straightforward to find that for the $\phi$
defined in the~\ref{sec:Tunneling-rate-for-complex}, the
classical minimal points will be placed on the axis of $u$, i.e.
$v=0$. In this case, the Fokker-Planck equation~(\ref{FPE_scaled})
is transformed to 
\begin{eqnarray}
\frac{\partial P}{\partial\tau} & = & \left\{ -\frac{\partial}{\partial u}\left[e^{i(\theta/2+\phi)}\cos\left(\Upsilon_{+}\right)\sin\left(\Upsilon_{-}\right)\right.\right.\nonumber \\
 &  & \hspace{2em}+e^{-i(\theta/2+\phi)}\cos\left(\Upsilon_{-}\right)\sin\left(\Upsilon_{+}\right)\nonumber \\
 &  & \left.\hspace{2em}-\bar{c}e^{i(\theta/2+\phi)}\tan\left(\Upsilon_{+}\right)-\bar{c}^{\ast}e^{-i(\theta/2+\phi)}\tan\left(\Upsilon_{-}\right)\right]\nonumber \\
 &  & -\frac{\partial}{\partial v}\left[e^{i(\theta/2-\phi)}\cos\left(\Upsilon_{+}\right)\sin\left(\Upsilon_{-}\right)\right.\nonumber \\
 &  & \hspace{2em}-e^{-i(\theta/2-\phi)}\cos\left(\Upsilon_{-}\right)\sin\left(\Upsilon_{+}\right)\nonumber \\
 &  & \left.\hspace{2em}-\bar{c}e^{i(\theta/2-\phi)}\tan\left(\Upsilon_{+}\right)+\bar{c}^{\ast}e^{-i(\theta/2-\phi)}\tan\left(\Upsilon_{-}\right)\right]\nonumber \\
 &  & \left.+\frac{\partial^{2}}{\partial u^{2}}\frac{\cos2\phi}{n}+\frac{\partial^{2}}{\partial v^{2}}\frac{\cos2\phi}{n}\right\} P.\label{eq:FPE_uv}
\end{eqnarray}
Here we have introduce another shifted coupling constant which interpolates
between the men-field coupling and the one used to analyse the $\beta$
manifold:
\begin{equation}
\bar{c}\equiv\tilde{c}+\frac{1}{2n},\label{eq:cbar}
\end{equation}
which effectively defines the relevant region of the phase diagram.
Thus the diffusion coefficients are all constant and equal, which
means that we can use the potential-barrier approximation~\cite{kramers1940brownian,landauer1961frequency,drummond1989quantum,kinsler1991quantum},
generalized in~\ref{sec:Tunneling-rate-for-complex}, to
obtain the tunneling rate.

We can obtain the Jacobean in the form of 
\begin{equation}
J=\left|\begin{array}{cc}
\frac{\partial\beta}{\partial u} & \frac{\partial\beta}{\partial v}\\
\frac{\partial\beta^{+}}{\partial u} & \frac{\partial\beta^{+}}{\partial v}
\end{array}\right|=-\frac{1}{2\cos2\phi}\cos\left[\Upsilon_{+}\right]\cos\left[\Upsilon_{-}\right].
\end{equation}
Using the relations $P'_{ss}(u,v)=JP_{ss}(\beta,\beta^{+})$ and $P'_{ss}(u,v)=N'\exp(-\Phi(u,v))$,
the potential is given by
\begin{eqnarray}
\Phi(u,v) & = & -n\left[2\sin\left[\Upsilon_{+}\right]\sin\left[\Upsilon_{-}\right]\right.\nonumber \\
 &  & \left.+\bar{c}\ln\left\{ \cos^{2}\left[\Upsilon_{+}\right]\right\} +\bar{c}^{\ast}\ln\left\{ \cos^{2}\left[\Upsilon_{-}\right]\right\} \right].\label{eq:potential_uv}
\end{eqnarray}
The minimal points of the potential in the $u$ and $v$ variables
are found where the gradient of the potential is zero. Thus we find
that the classical minimal points are located at $(u^{(c)},v^{(c)})=(\pm2r\cos2\phi,0)$,
where 
\begin{equation}
re^{i\psi}\equiv\sin^{-1}B=\sin^{-1}\left[1+\left(\bar{c}^{2}-|\bar{c}|^{2}\right)/2-\bar{c}\Pi(\bar{c})\right]^{1/2},\label{minima1}
\end{equation}
with $\phi=\psi-\theta/2$ and hence 
\begin{equation}
re^{-i\psi}=\sin^{-1}B^{\ast}=\sin^{-1}\left[1+\left(\bar{c}^{*2}-|\bar{c}|^{2}\right)/2-\bar{c}^{*}\Pi(\bar{c})\right]^{1/2}.\label{minima2}
\end{equation}
From the manifold~(\ref{eq:manifold}) and the transformation~(\ref{eq:transform}),
we will find that the line of $v=0$ with real $u$ is on this manifold,
where we will have $\Upsilon_{+}=\Upsilon_{-}^{\ast}$. Hence the
potential~(\ref{eq:potential_uv}) is proved to be real. In the meanwhile,
the classical minimal points and the saddle point at the origin are
all on this line. Considering 
\begin{equation}
\frac{\partial P_{1}}{\partial\tau}+\nabla\cdot\vec{J}=0,
\end{equation}
the current $\vec{J}$ can be obtained easily via the Fokker-Planck
equation~(\ref{eq:FPE_uv}). It is directly checked that the current
through this line $J_{u}$ is real. This shows that quantum tunneling
mostly occurs through this line. Further analysis is given in the~\ref{sec:Tunneling-rate-for-complex}.

The second derivatives on this line are always real as well, given
that 
\begin{eqnarray}
\Phi_{uu} & = & \frac{n}{\cos^{2}(2\phi)}\left[\cos(\theta+2\phi)\sin(\Upsilon_{+})\sin(\Upsilon_{-})\right.\nonumber \\
 &  & \left.-\cos(\Upsilon_{+})\cos(\Upsilon_{-})+\frac{\bar{c}\exp[i(\theta+2\phi)]}{2\cos^{2}(\Upsilon_{+})}+\frac{\bar{c}^{\ast}\exp[-i(\theta+2\phi)]}{2\cos^{2}(\Upsilon_{-})}\right],\nonumber \\
\Phi_{vv} & = & \frac{n}{\cos^{2}(2\phi)}\left[\cos(\theta-2\phi)\sin(\Upsilon_{+})\sin(\Upsilon_{-})\right.\nonumber \\
 &  & \left.+\cos(\Upsilon_{+})\cos(\Upsilon_{-})+\frac{\bar{c}\exp[i(\theta-2\phi)]}{2\cos^{2}(\Upsilon_{+})}+\frac{\bar{c}^{\ast}\exp[-i(\theta-2\phi)]}{2\cos^{2}(\Upsilon_{-})}\right].\nonumber \\
\label{eq:sec-der-positive}
\end{eqnarray}
Therefore, the potentials of the saddle points and the classical minimal
points are
\begin{eqnarray}
\Phi^{(o)} & = & 0,\nonumber \\
\Phi^{(c)} & = & -n\left[2|B|^{2}+\bar{c}\ln(1-B^{2})+\bar{c}^{\ast}\ln(1-B^{\ast2})\right].\label{eq:potential-barrier}
\end{eqnarray}

The related second derivatives are therefore
\begin{eqnarray}
\Phi_{uu}^{(o)} & \equiv & \frac{\partial^{2}\Phi}{\partial u^{2}}(0,0)=n\left(\frac{-2+\bar{c}\exp[i(\theta+2\phi)]+\bar{c}^{\ast}\exp[-i(\theta+2\phi)]}{2\cos^{2}(2\phi)}\right),\nonumber \\
\Phi_{vv}^{(o)} & \equiv & \frac{\partial^{2}\Phi}{\partial v^{2}}(0,0)=n\left(\frac{2+\bar{c}\exp[i(\theta-2\phi)]+\bar{c}^{\ast}\exp[-i(\theta-2\phi)]}{2\cos^{2}(2\phi)}\right),\nonumber \\
\Phi_{uu}^{(c)} & = & \frac{n}{2\cos^{2}(2\phi)}\left[-2\sqrt{1-B^{2}}\sqrt{1-B^{\ast2}}+2\cos(\theta+2\phi)|B|^{2}\right.\nonumber \\
 &  & \left.+\frac{\bar{c}\exp[i(\theta+2\phi)]}{1-B^{2}}+\frac{\bar{c}^{\ast}\exp[-i(\theta+2\phi)]}{1-B^{\ast2}}\right],\nonumber \\
\Phi_{vv}^{(c)} & = & \frac{n}{2\cos^{2}(2\phi)}\left[2\sqrt{1-B^{2}}\sqrt{1-B^{\ast2}}+2\cos(\theta-2\phi)|B|^{2}\right.\nonumber \\
 &  & \left.+\frac{\bar{c}\exp[i(\theta-2\phi)]}{1-B^{2}}+\frac{\bar{c}^{\ast}\exp[-i(\theta-2\phi)]}{1-B^{\ast2}}\right].\label{eq:sec-der-pb}
\end{eqnarray}
The tunneling time for a symmetric bistable potential in two dimensions
is calculated using an extension of the Kramers method developed by
Landauer and Swanson~\cite{kramers1940brownian,landauer1961frequency},
which is called the potential-barrier approximation~\cite{drummond1989quantum,kinsler1991quantum}
and generalized for complex cases in~\ref{sec:Tunneling-rate-for-complex}.

In the~\ref{sec:Tunneling-rate-for-complex}, we have obtained
the analytic formation of the tunneling time as shown in equation~(\ref{eq:tunneling_complex}).
Since the potential and the related second derivatives have been calculated
in equations~(\ref{eq:potential-barrier}) and (\ref{eq:sec-der-pb}),
the analytic result of the tunneling time is:
\begin{equation}
T=\frac{2\pi}{|g|\cos2\phi}\left[\frac{-\Phi_{vv}^{(o)}}{\Phi_{uu}^{(o)}\Phi_{uu}^{(c)}\Phi_{vv}^{(c)}}\right]^{\frac{1}{2}}\exp(\Phi^{(o)}-\Phi^{(c)})\,,
\end{equation}
which can be expressed in terms of our parameters as:
\begin{eqnarray}
T & = & \frac{4\pi\cos2\phi}{\mathcal{E}}\exp(n\left[2|B|^{2}+\bar{c}\ln(1-B^{2})+\bar{c}^{*}\ln(1-B^{*2})\right])\nonumber \\
 &  & \times\left(\frac{\bar{c}e^{i(\theta-2\phi)}}{1-B^{2}}+\frac{\bar{c}^{\ast}e^{-i(\theta-2\phi)}}{1-B^{\ast2}}+2|\bar{c}|+2|B|^{2}\cos(\theta-2\phi)\right)^{-1/2}\nonumber \\
 &  & \times\left(\frac{\bar{c}e^{i(\theta+2\phi)}}{1-B^{2}}+\frac{\bar{c}^{\ast}e^{-i(\theta+2\phi)}}{1-B^{\ast2}}-2|\bar{c}|+2|B|^{2}\cos(\theta+2\phi)\right)^{-1/2}\nonumber \\
 &  & \times(2+\bar{c}e^{i(\theta-2\phi)}+\bar{c}^{\ast}e^{-i(\theta-2\phi)})^{1/2}\nonumber \\
 &  & \times(2-\bar{c}e^{i(\theta+2\phi)}-\bar{c}^{\ast}e^{-i(\theta+2\phi)})^{-1/2}.\label{tunneling_time}
\end{eqnarray}
Here we have used the relation $(1-B^{2})(1-B^{\ast2})=|\bar{c}|^{2}$,
which can be checked via equations~\eref{minima1} and~\eref{minima2}.
If we set $\theta=\phi=0$ and $\bar{c}=\bar{c}^{\ast}$, which means
all the parameters are real, the tunneling time~(\ref{tunneling_time})
can be simplified to give
\begin{equation}
T=\frac{\pi\sqrt{1+\bar{c}}}{\mathcal{E}(1-\bar{c})}\exp\left\{ 2n\left[1-\bar{c}+\bar{c}\ln\left(\bar{c}\right)\right]\right\} .
\end{equation}
This simplified form agrees with equation~(4.22) of Ref.~\cite{kinsler1991quantum},
with parameters are defined as $n=\mu/g^{2}$, $\bar{c}=\bar{\sigma}/\mu$
and $g=g^{2}\gamma_{1}$. Thus, in the situation where all the parameters
are real numbers, without anharmonic terms or detunings, the tunneling
time~(\ref{tunneling_time}) reduces to previous results Ref.~\cite{drummond1989quantum,kinsler1991quantum}.

\section{Number-state calculations\label{sec:Number-state-calculations}}

The tunneling time can also be obtained by solving the master equation~(\ref{eq:master_eq})
numerically in the number-state basis. In this basis the master equation
reduces to an infinite matrix equation. Nevertheless, as any physical
system has a finite energy, a suitable energy cutoff will reduce the
system to a finite matrix equation. While this method is only numerically
feasible for small photon number, it allows us to check the accuracy
of the approximate analytic calculation given above.

\subsection{Number-state basis expansions}

We first expand the density operator $\rho$ in terms of its number-state
matrix elements $\rho_{kl}$, which are defined by 
\begin{equation}
\rho_{kl}=\langle k|\rho|l\rangle.
\end{equation}
Thus, the time evolution is given by 
\begin{equation}
\frac{d}{dt}\rho_{ij}=\left\langle i\left|\frac{d}{dt}\rho\right|j\right\rangle .
\end{equation}
Then the master equation~(\ref{eq:master_eq}) under number-state
basis becomes 
\begin{equation}
\frac{d}{dt}\rho_{ij}=T_{ij}^{kl}\rho_{kl}.
\end{equation}
Here we have used the Einstein summation convention on identical indices.
And $T_{ij}^{kl}$ is a four-dimensional transition matrix describing
the rate of transition from the state $\rho_{kl}$ to the state $\rho_{ij}$,
which is in the form of 
\begin{eqnarray}
T_{ij}^{kl} & = & \frac{\mathcal{E}}{2}\sqrt{i(i-1)}\delta_{i;j}^{k+2;l}-\frac{\mathcal{E}}{2}\sqrt{(j+1)(j+2)}\delta_{i;j}^{k;l-2}\nonumber \\
 &  & +\frac{\mathcal{E}}{2}\sqrt{j(j-1)}\delta_{i;j}^{k;l+2}-\frac{\mathcal{E}}{2}\sqrt{(i+1)(i+2)}\delta_{i;j}^{k-2;l}\nonumber \\
 &  & -\left[\gamma i+\gamma^{\ast}j+\frac{g}{2}i(i-1)+\frac{g^{\ast}}{2}j(j-1)\right]\delta_{i;j}^{k;l}\nonumber \\
 &  & +\gamma^{(2)}\sqrt{(i+1)(i+2)(j+1)(j+2)}\delta_{i;j}^{k-2;l-2}\nonumber \\
 &  & +2\gamma_{1}^{(1)}\sqrt{(i+1)(j+1)}\delta_{i;j}^{k-1;l-1}.
\end{eqnarray}
Here we have used the assumption $\Delta_{2}=0$ so that $\mathcal{E}$
is real, and 
\begin{equation}
	\delta_{i;j}^{k;l}=\cases{1 & if $i=k$ and $j=l$,\\	0 & otherwise.}
\end{equation}
The behavior of the system can be characterized in terms of the eigenvalues
and eigenvectors of the transition matrix $T_{ij}^{kl}$. For instance,
the eigenvector corresponding to the zero eigenvalue is exactly the
steady state of the system. The first negative eigenvalue is related
to the quantum tunneling rate~\cite{kinsler1991quantum}.

In order to make the matrix finite, we set a photon number cutoff
$N$ so that $0\le i,j,k,l\le N$. This approximation is valid if
the high-photon-number states that are ignored play no significant
role in determining the evolution of the system. The four-dimensional
matrix $T_{ij}^{kl}$ can be reduced to a two-dimensional one $T_{\bar{\alpha}}^{\bar{\beta}}$
with this truncation that 
\begin{equation}
\frac{d}{dt}\rho_{\bar{\alpha}}=T_{\bar{\alpha}}^{\bar{\beta}}\rho_{\bar{\beta}},
\end{equation}
with 
\begin{eqnarray}
T_{\bar{\alpha}}^{\bar{\beta}} & = & \frac{\mathcal{E}}{2}\sqrt{i(i-1)}\delta_{\bar{\alpha}}^{\bar{\beta}+2N+2}-\frac{\mathcal{E}}{2}\sqrt{(j+1)(j+2)}\delta_{\bar{\alpha}}^{\bar{\beta}-2}\nonumber \\
 &  & +\frac{\mathcal{E}}{2}\sqrt{j(j-1)}\delta_{\bar{\alpha}}^{\bar{\beta}+2}-\frac{\mathcal{E}}{2}\sqrt{(i+1)(i+2)}\delta_{\bar{\alpha}}^{\bar{\beta}-2N-2}\nonumber \\
 &  & -\left[\gamma i+\gamma^{\ast}j+\frac{g}{2}i(i-1)+\frac{g^{\ast}}{2}j(j-1)\right]\delta_{\bar{\alpha}}^{\bar{\beta}}\nonumber \\
 &  & +\gamma^{(2)}\sqrt{(i+1)(i+2)(j+1)(j+2)}\delta_{\bar{\alpha}}^{\bar{\beta}-2N-4}\nonumber \\
 &  & +2\gamma_{1}^{(1)}\sqrt{(i+1)(j+1)}\delta_{\bar{\alpha}}^{\bar{\beta}-N-2},\label{eq:transition-matrix}
\end{eqnarray}
where 
\begin{eqnarray}
\bar{\alpha} & = & (N+1)i+j+1,\quad\bar{\beta}=(N+1)k+l+1.
\end{eqnarray}
Here $\delta_{\bar{\alpha}}^{\bar{\beta}}$ is a Kronecker delta,
and $\bar{\alpha}$, $\bar{\beta}$ are in the range of $[1,(N+1)^{2}]$.
Note that the transition matrix is not Hermitian because of the single-
and two-photon decay process.

We label the $k$-th eigenvalue by $\epsilon_{k}$ and its corresponding
eigenvector by $\rho_{\bar{\alpha}}^{(k)}$ so that 
\begin{equation}
\rho_{\bar{\alpha}}(t)=\sum_{k\ge0}A_{k}\exp(\epsilon_{k}t)\rho_{\bar{\alpha}}^{(k)}.
\end{equation}
Here the coefficients $A_{k}$ define the initial state. We order
the indices $k$ by the size of the real part of the eigenvalues,
so that $\Re(\epsilon_{k})\ge\Re(\epsilon_{k+1})$. $\epsilon_{0}$
is the stable eigenvalue ($\epsilon_{0}=0$) and $\rho_{\bar{\alpha}}^{(0)}$
is the stable state. $\epsilon_{1}$ is the tunneling eigenvalue so
the tunneling time is obtained~\cite{kinsler1991quantum}, 
\begin{equation}
T_{N}=-\frac{2}{\epsilon_{1}}.\label{tunneling_time_number}
\end{equation}

\begin{figure}
\includegraphics[width=0.48\textwidth]{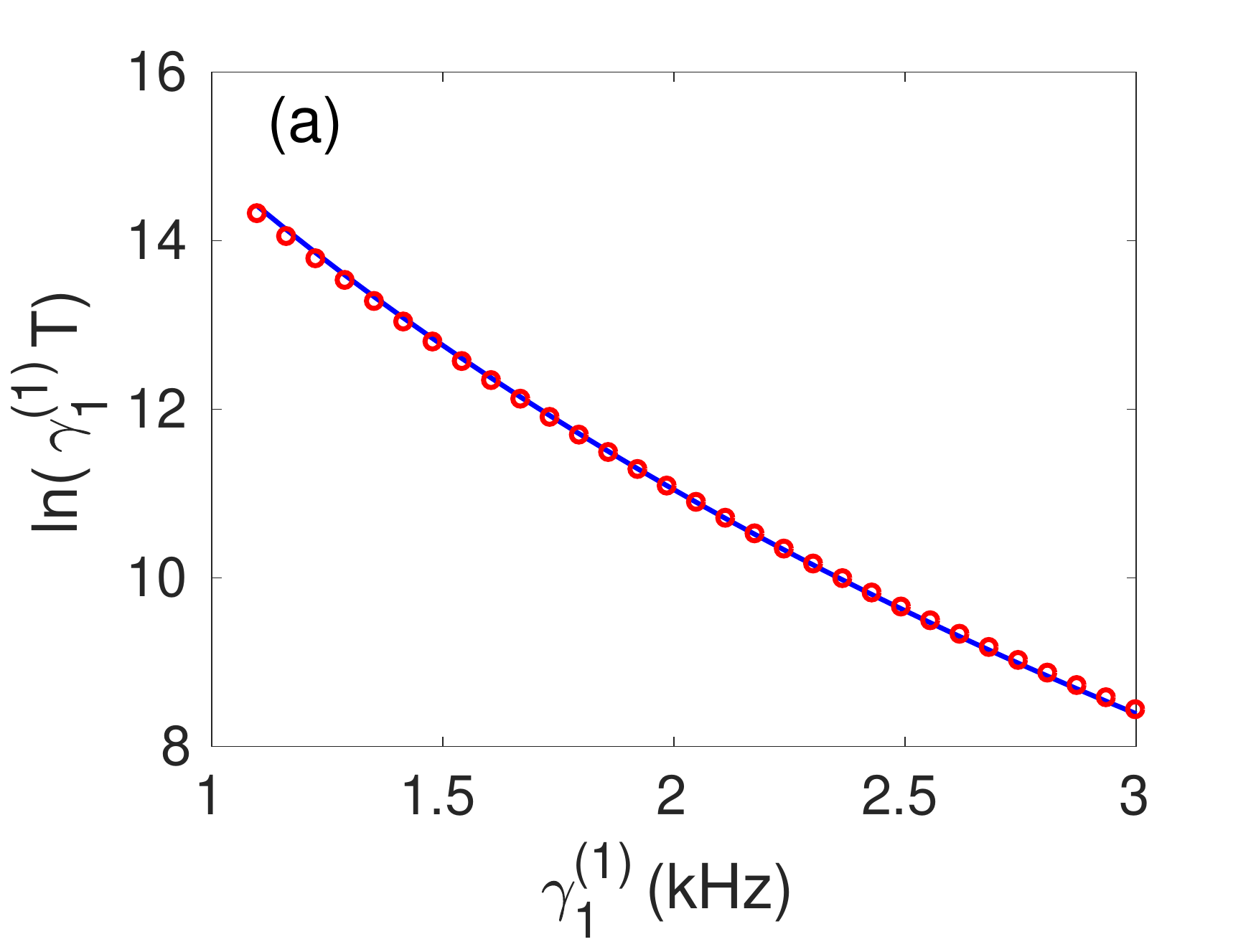} \includegraphics[width=0.48\textwidth]{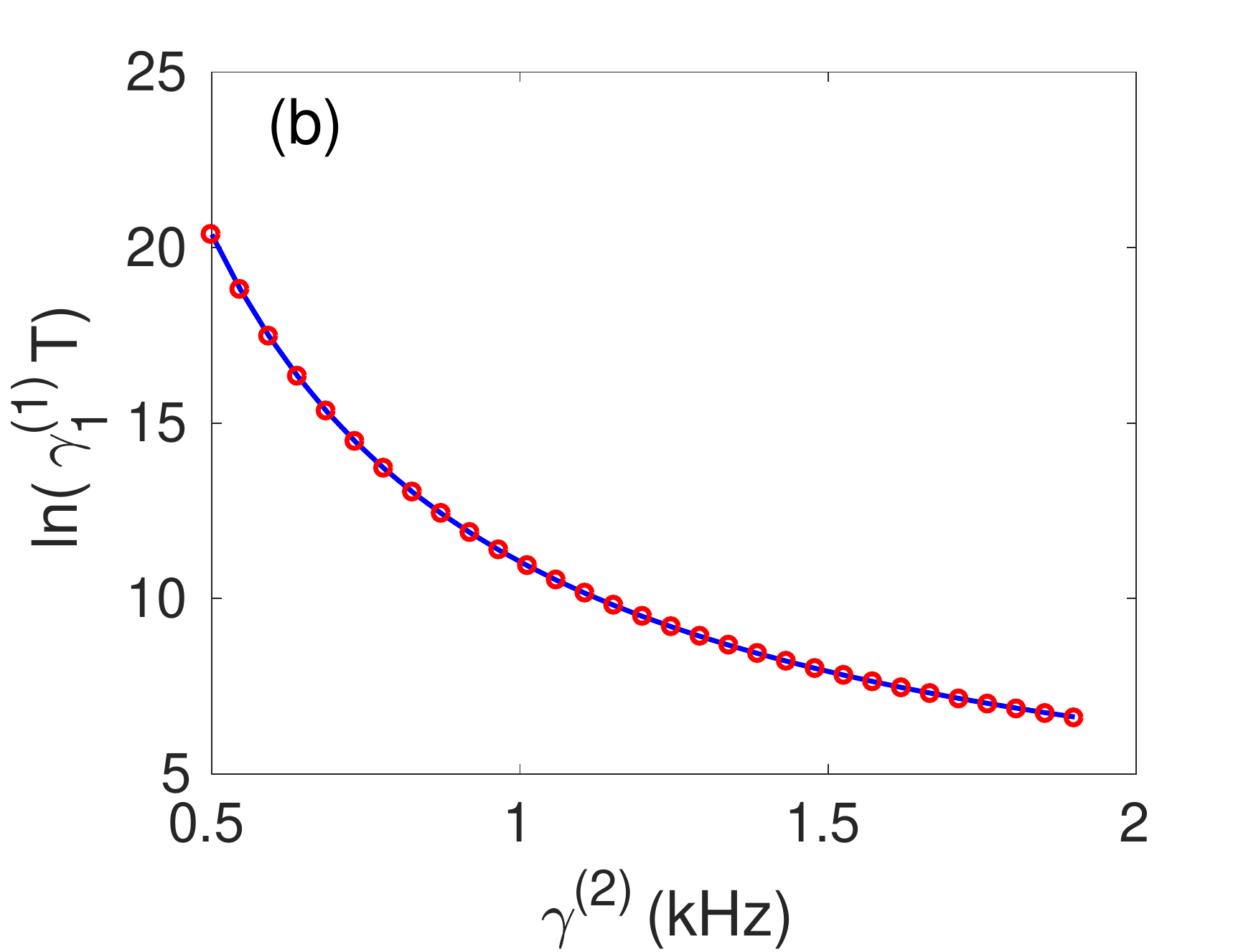}\\
 \includegraphics[width=0.48\textwidth]{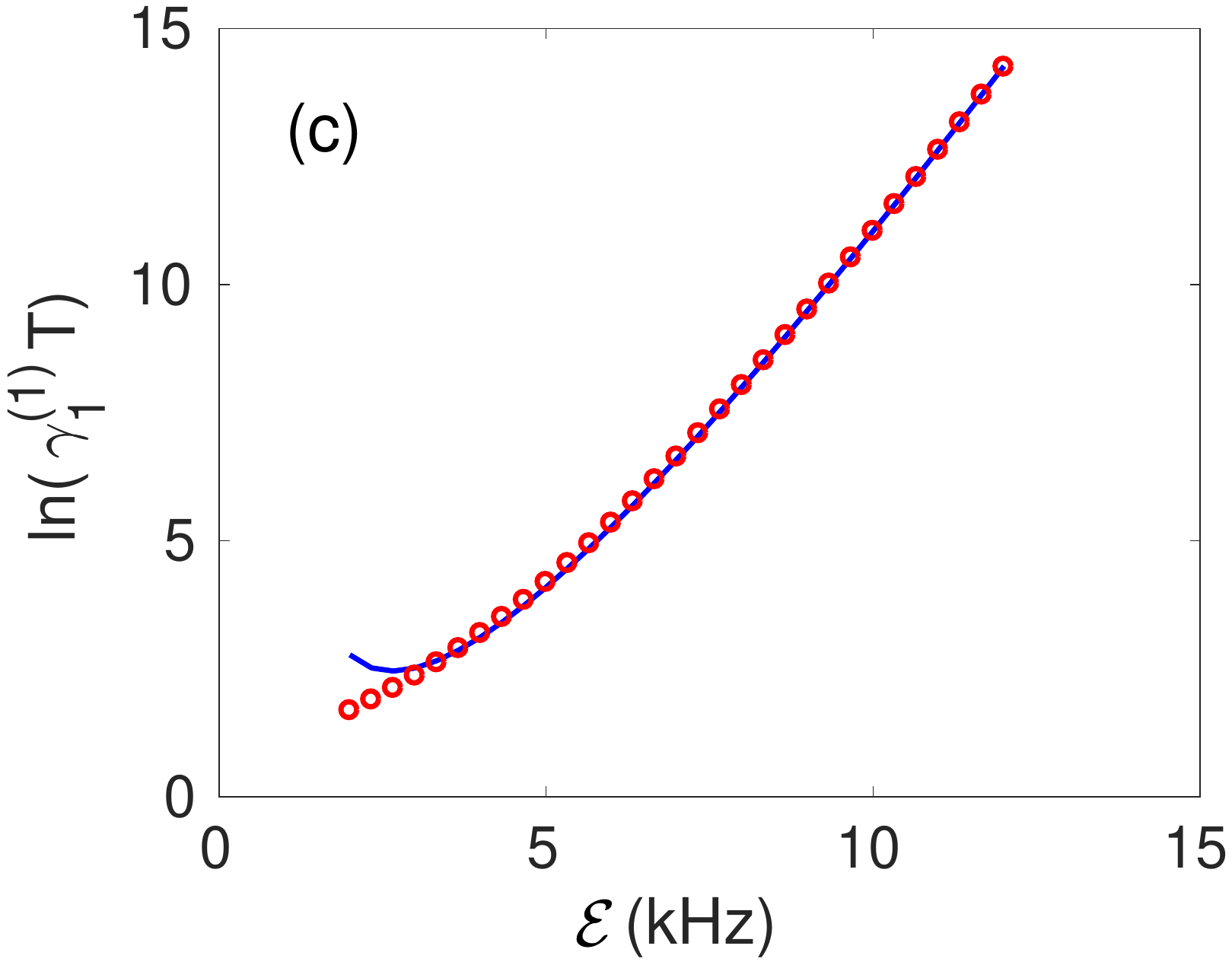} \includegraphics[width=0.48\textwidth]{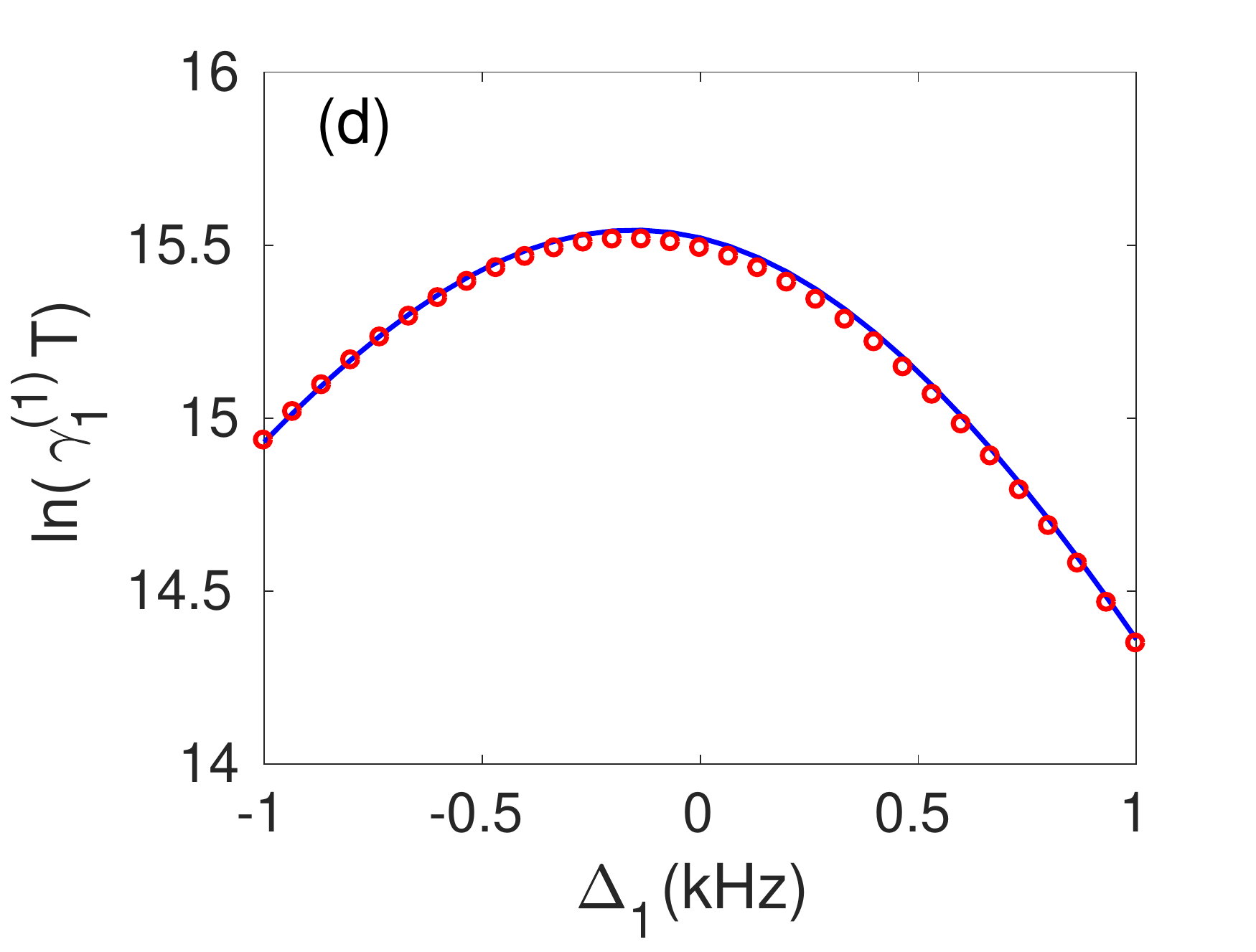}
\caption{Comparisons of the tunneling time obtained by P-representation~(\ref{tunneling_time})
(blue lines) and those by number-state expansion~(\ref{tunneling_time_number})
(red circles) changing with $\gamma_{1}^{(1)}$ (a), $\gamma^{(2)}$
(b), $\mathcal{E}$ (c), and $\Delta_{1}$ (d), respectively. In the
figure (a), other parameters are $\Delta_{1}=0$, $\gamma^{(2)}=1$kHz,
$\chi=0.1$kHz, $\mathcal{E}=10$kHz, and $n=9.95$. In the figure
(b), $\gamma=2$kHz, $\chi=0.1$kHz, $\mathcal{E}=10$kHz. In the
figure (c), $\gamma=2$kHz, $\gamma^{(2)}=1$kHz, $\chi=0.1$kHz.
And in the figure (d), $\gamma_{1}^{(1)}=1.5$kHz, $\gamma^{(2)}=0.8$kHz,
$\chi=0.1$kHz, $\mathcal{E}=10$kHz, and $n=12.40$. We note that
$\tilde{c}=(\gamma-g)/\left(gn\right)$, $\gamma=\gamma_{1}^{(1)}+i\Delta_{1}$,
$n=|\mathcal{E}/g|$, and $g=\gamma^{(2)}+i\chi$, so the parameters
$\tilde{c}$ and $n$ will change with $\gamma_{1}^{(1)}$ (a), $\gamma^{(2)}$
(b), $\mathcal{E}$ (c) and $\Delta_{1}$ (d). In all the figures,
the number-state expansion results are obtained with a particle number
cut-off $N=70$, after the adiabatic elimination.}
\label{fig:log-time-chi}
\end{figure}

\subsection{Tunneling time calculations and comparisons}

We will compare the tunneling times obtained from using the P-representation~(\ref{tunneling_time})
with those obtained using a number-state expansion~(\ref{tunneling_time_number}).
By changing the parameters $\gamma_{1}^{(1)}$, $\gamma^{(2)}$, $\mathcal{E}$
and $\Delta_{1}$, the results for the tunneling time are shown in
the figure~\ref{fig:log-time-chi}. Here we have noted that parameter
$\tilde{c}$ consists of $\gamma_{1}^{(1)}$, $\gamma^{(2)}$, $\chi$,
$\Delta_{1}$ and $\mathcal{E}$ as shown in equation~\eref{eq:c},
and $n$ consists of $\gamma^{(2)}$, $\chi$ and $\mathcal{E}$ shown
in equation~(\ref{eq:parameter}). The parameters has been chosen
to satisfy $\Re(\tilde{c})>0$, with large single-photon loss and
small nonlinear couplings, and $|\tilde{c}|<1$, above threshold.

Since there is a potential barrier between the minima, quantum tunneling
is a slow effect. Hence, we use $\ln(\gamma_{1}^{(1)}T)$ to compare
the results in the limit of large tunneling time $T$. It is expected
that the analytic potential-barrier approximation will be most reliable
for large tunneling times. We show below that, as expected, the analytic
results agree exceptionally well with numerical results for large
tunneling times, which is the limit of most interest for understanding
spontaneously broken symmetry.

Figures~\ref{fig:log-time-chi}~(a) and (b) show that damping speeds
up quantum tunneling. Figure~\ref{fig:log-time-chi}~(c) shows that
an increased driving will increase the tunneling time, since the parameter
$\mathcal{E}$ is proportional to the driving $\mathcal{E}_{2}$.
The change in tunneling time with the detuning $\Delta_{1}$ is shown
in figure~\ref{fig:log-time-chi}~(d), where the largest tunneling
time occurs at $\Delta_{1}=-0.2$kHz because of the nonzero nonlinearity
$\chi$. With large detuning $\Delta_{1}$, the tunneling time will
be reduced as shown in the figure~\ref{fig:log-time-chi}~(d). It
is also clear in the figure~\ref{fig:log-time-chi}~(c) that the
potential-barrier approximation fails when the quantum tunneling becomes
too fast, as expected for this approach. 

The behaviors of tunneling time changing with driving can be understood
from the analytic equations. At large driving $\mathcal{E}$, the
parameter $n$ increases while $n\bar{c}$ remains unchanged. From
the formation of the classical stable points~(\ref{minimal_point}),
it is directly checked that the stable points move away from each
other in this case. In addition, the potential barrier becomes larger
according to equations~(\ref{eq:potential-barrier}). Therefore,
the quantum tunneling is suppressed when we increase the driving $\mathcal{E}$
considering the analytic result~(\ref{tunneling_time}) provides
$T\sim\exp(\Phi^{(o)}-\Phi^{(c)})$.

A similar analysis can be applied on the results of figure~\ref{fig:log-time-chi}~(a)
and (b) as well. When the dampings $\gamma_{1}^{(1)}$ and $\gamma^{(2)}$
become larger, we find that the positions of the stable points~(\ref{minimal_point})
only change slightly, but the potential barrier~(\ref{eq:potential-barrier})
will greatly decrease. Thus, increasing the dampings $\gamma_{1}^{(1)}$
and $\gamma^{(2)}$ speed up quantum tunneling.

\begin{figure}
\centering \includegraphics[width=0.48\textwidth]{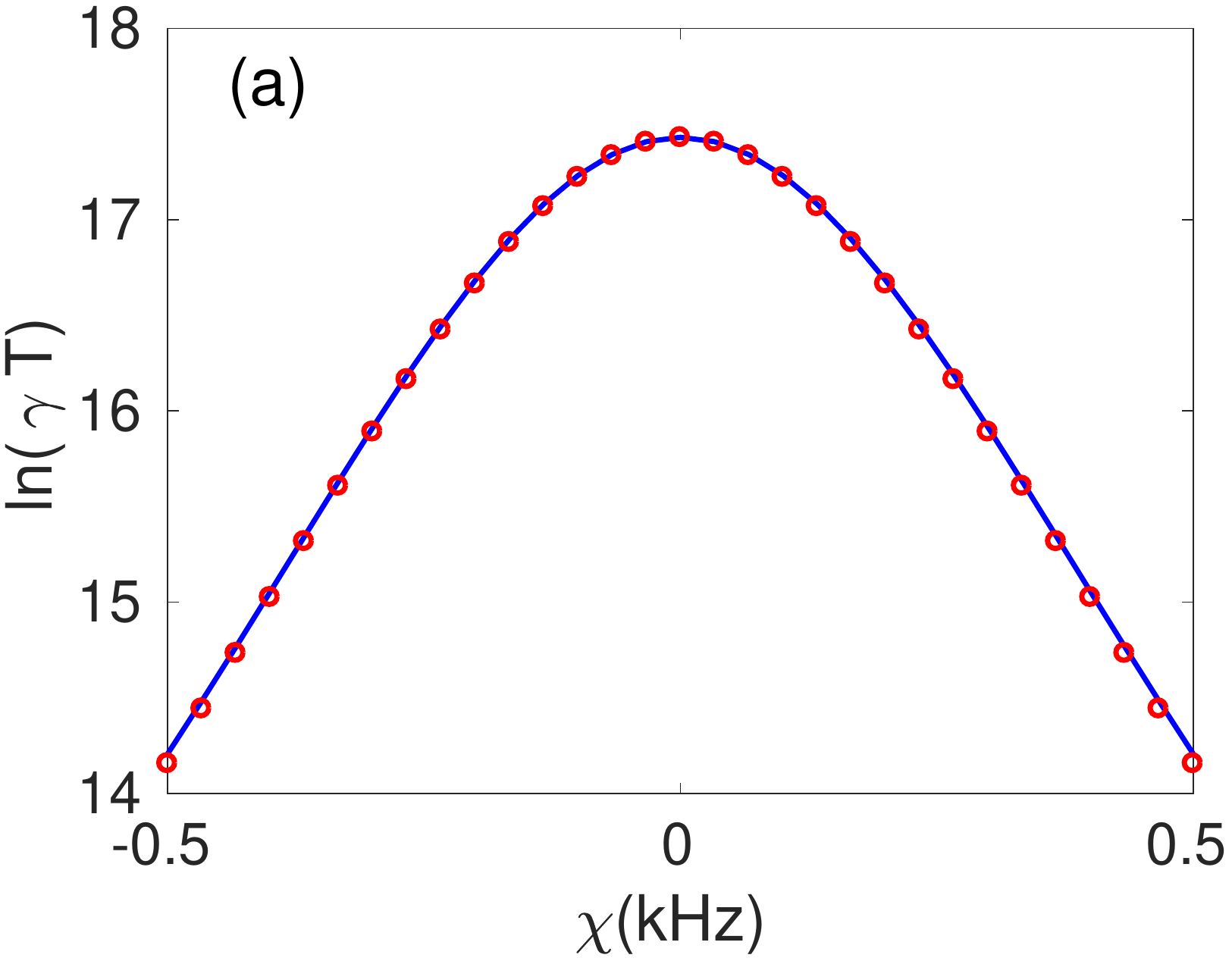} \includegraphics[width=0.48\textwidth]{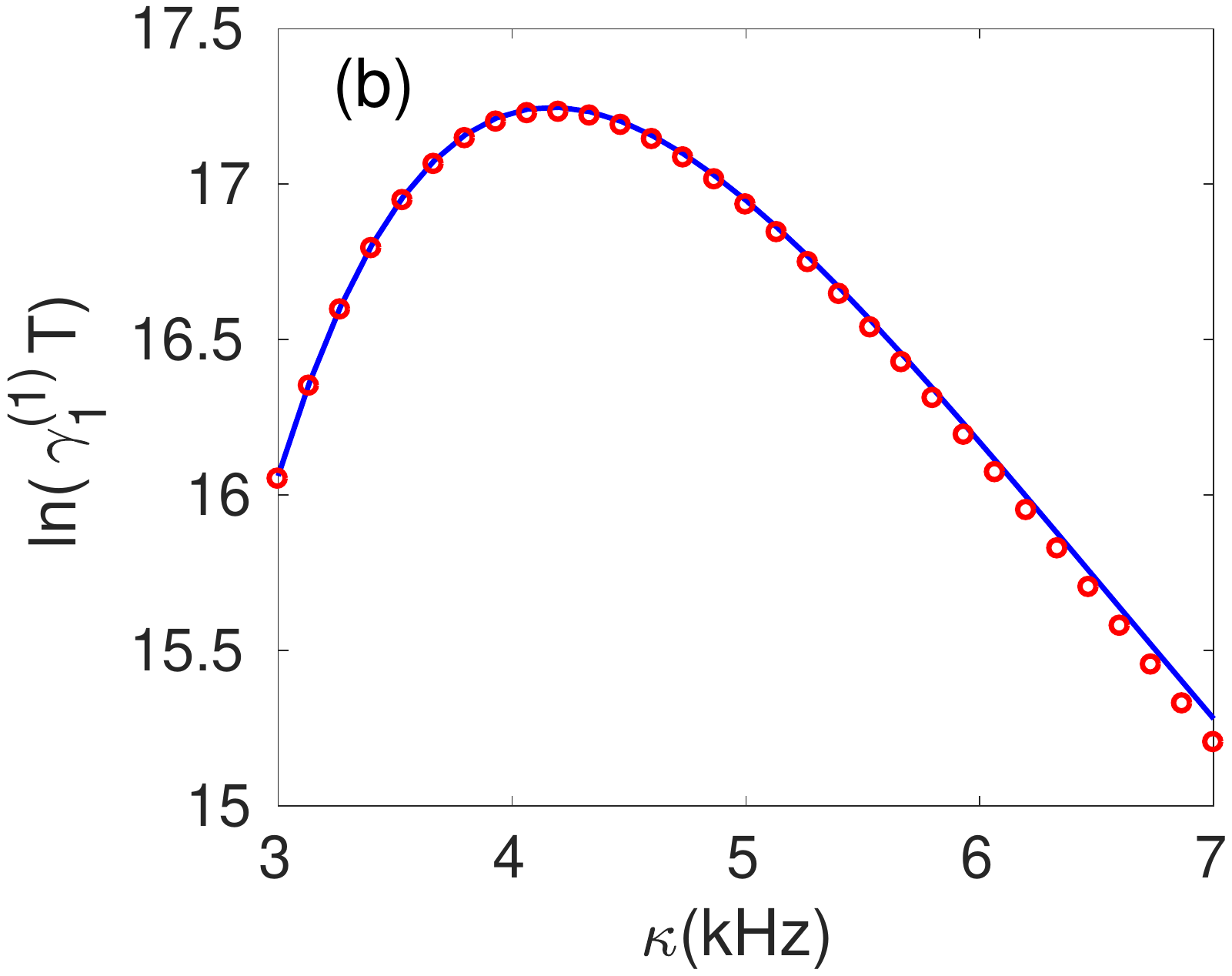}
\caption{Comparisons of the tunneling times obtained using the P-representation~(\ref{tunneling_time})
(blue lines) and those using the number-state expansion~(\ref{tunneling_time_number})
(red circles) changing with the anharmonic nonlinearity $\chi$ (a)
and the parametric nonlinearity $\kappa$ (b). In figure (a), the
other parameters are $\gamma=1.5$kHz, $\gamma^{(2)}=0.5$kHz, $\mathcal{E}=8$kHz.
In figure (b), $\gamma=1.5$kHz, $\gamma_{1}^{(2)}=0.1$kHz, $\gamma_{2}=20$kHz,
$\chi=0.1$kHz, $\mathcal{E}_{2}=40$kHz. Because $\tilde{c}=(\gamma-g)/\left(gn\right)$,
$\gamma=\gamma_{1}^{(1)}+i\Delta_{1}$, $g=\gamma^{(2)}+i\chi=\gamma_{1}^{(2)}+\kappa^{2}/(2\gamma_{2})+i\chi$,
$n=|\mathcal{E}/g|$ and $\mathcal{E}=\kappa\mathcal{E}_{2}/\gamma_{2}$,
the parameters $\tilde{c}$ and $n$ will change with the $\chi$
(a) and $\kappa$ (b). In both figures, number-state expansion results
are obtained with a particle number cut-off $N=100$, after adiabatic
elimination.}
\label{fig:log-time-nonlinearity}
\end{figure}

The dependence of the tunneling rate changing on the nonlinearities
$\chi$ and $\kappa$ are shown in figure~\ref{fig:log-time-nonlinearity}.
From the formations~(\ref{eq:c}) and (\ref{eq:cbar}), $\tilde{c}$
and $\bar{c}$ will be real if $\chi=\gamma^{(2)}\Delta_{1}/\gamma_{1}^{(1)}$
considering there is always a nonzero damping $\gamma_{1}^{(1)}$
in realistic systems. In the case of $\Delta_{1}=0$, we will have
$\chi=0$ when $\bar{c}$ is real. This is the case shown in figure~\ref{fig:log-time-nonlinearity}~(a),
where $\chi=0$ corresponds to the largest tunneling time. This shows
that the nonlinear coupling $\chi$ will decrease the tunneling time.
This result is consistent with the previous ones in similar by simpler
systems~\cite{Casteels_PRA2016,Rodriguez_PRL2017}.

This behavior can also be obtained from the analytic result for the
tunneling time~(\ref{tunneling_time}). Considering the definitions
$\bar{c}=(2\gamma-g)/(2gn)$, $n=\left|\mathcal{E}/g\right|$ and
$g=\gamma^{(2)}+i\chi$, it is directly checked that $\bar{c}$ and
$n$ will decrease if the nonlinearity $\chi$ becomes larger. Then
from the results of the potential for the saddle point and the classical
stable points~(\ref{eq:potential-barrier}), we see that the difference
of the potentials becomes smaller. Hence, the nonlinearity $\chi$
will reduce the potential barrier height and thus speed up the tunneling.

The effects of the nonlinearity $\kappa$ are more complicated than
$\chi$ as shown in figure~\ref{fig:log-time-nonlinearity}~(b),
which were not treated in previous study on single-mode nonlinear
resonators~\cite{Casteels_PRA2016,Rodriguez_PRL2017}. Here we show
that $\kappa$ will increase the tunneling time when it is small,
and decrease the tunneling time once it is large enough. This can
be understood by considering that $\mathcal{E}\propto\kappa$ and
$\gamma^{(2)}\propto\kappa^{2}$ if $\gamma_{1}^{(2)}$ is negligible.
Thus when $\kappa$ is small, the effect of $\mathcal{E}$ dominates,
which will increase the tunneling time. When $\kappa$ is large enough
so that the effect of $\gamma^{(2)}$ becomes more important, the
quantum tunneling will then be sped up.

We also note that the analytical results are not expected to agree
very well with the numerical results with large nonlinearities $\chi$
and $\kappa$ if other parameters are fixed. This is because the dimensionless
driving field is smaller, which also decreases the potential barrier
height, and therefore reduces the validity of the analytic approximations
used for tunneling calculations. Essentially, in this limit there
is no real tunneling, and the broken time-translation symmetry is
rapidly restored.

\section{Conclusion}

\label{sec:Conclusions}

In this paper, we have studied general quantum subharmonic generation
with additional detunings and anharmonicity, which has been experimental
achieved~\cite{leghtas2015confining} in superconducting microwave
cavities. With driving, damping and nonlinearity considered, we obtained
the steady-state solution of the Fokker-Planck equation using the
adiabatic approximation and the zero-temperature limit, in order to
understand pure quantum tunneling effects. Because of the nonlinearity,
a complex parameter $\tilde{c}$ has been introduced. This means that
the potential of the steady state is in general complex, which is
different from the previously studied quantum optical subharmonic
generation systems where the potentials were always real.

Quantum tunneling has been studied in this non-equilibrium system.
This is related to quantum time symmetry breaking, as it defines the
maximum time that a spontaneously broken time phase can exist before
randomly switching to a different discrete time phase. By studying
the manifold of the steady-state potential, we find that quantum tunneling
will occur in the parameter region of $\Re(\tilde{c})>0$ and $|\tilde{c}|<1$,
i.e the region of large single-photon loss, small nonlinear couplings
and above threshold. The tunneling time has been obtained analytically
using the potential-barrier approximation. In the expected domain
of applicability of large tunneling time, the results agree with numerical
calculations using a number-state basis.

These results show that the anharmonicity $\chi$ will enhances quantum
tunneling rates compared to previous cases with no anharmonic term.
This may have practical applications for escaping a local minimum
in quantum neural networks \cite{inagaki2016coherent,mcmahon2016fully},
where the global potential minimum is the desired computational solution.

\ack This work is supported by the National Key R\&D Program of China
(Grants No. 2016YFA0301302 and No. 2018YFB1107200), the National Natural
Science Foundation of China (Grants No. 11622428, No. 61475006, and
No. 61675007) and the Graduate Academic Exchange Fund of Peking University.
PDD and MDR thank the Australian Research Council and the hospitality
of the Institute for Atomic and Molecular Physics (ITAMP) at Harvard
University, supported by the NSF. This research has also been supported
by the Australian Research Council Discovery Project Grants schemes
under Grants DP180102470 and DP190101480.

\appendix

\section{Tunneling rate for complex Fokker-Planck equations\label{sec:Tunneling-rate-for-complex}}

Previous research on tunneling rates~\cite{landauer1961frequency,kinsler1991quantum,drummond1989quantum,kramers1940brownian}
has treated real potentials. For our system, we obtain a complex potential
barrier and a complex Fokker-Planck equation. Thus, we must generalize
the potential-barrier approximation~\cite{landauer1961frequency,kramers1940brownian}
to treat such complex Fokker-Planck cases.

Without loss of generality, we will study the complex Fokker-Planck
equation 
\begin{eqnarray}
\frac{\partial P}{\partial\tau} & = & \left[e^{i\theta}\left(\frac{\partial}{\partial\zeta}A(\zeta,\zeta^{+})+\frac{1}{2n}\frac{\partial^{2}}{\partial\zeta^{2}}\right)+h.c.\right]P.
\end{eqnarray}
The notation $h.c.$ has the same meaning as we introduced in \sref{sec:Adiabatic-Elimination}.
Our Fokker-Planck equation~(\ref{FPE_scaled}) can be written in
this form with the transformation 
\begin{equation}
\zeta=\sin^{-1}\beta,\quad\zeta^{+}=\sin^{-1}\beta^{+}.\label{eq:transform-1}
\end{equation}
The manifold we are concerned with~(\ref{eq:manifold}) is then transformed
into:
\begin{eqnarray}
\zeta & =\sin^{-1}\left[x+ix\tan\left(\varphi\right)\cos^{p}\left(x\pi/2\right)\cos^{p}\left(y\pi/2\right)\right],\nonumber \\
\zeta^{+} & =\sin^{-1}\left[y-iy\tan\left(\varphi\right)\cos^{p}\left(x\pi/2\right)\cos^{p}\left(y\pi/2\right)\right],\label{eq:manifold-1}
\end{eqnarray}
with the classical stationary points $\beta^{(c)}=\pm e^{i\varphi}\left|\beta^{(c)}\right|$
transformed into $\zeta^{(c)}=\pm re^{i\psi}$. This manifold gives
the correct behavior in the limit of $p\rightarrow0$, as introduced
in the section~\ref{subsec:Complex-phase-space-manifold}. The steady-state
solution can be expressed by a potential $\Phi$, which satisfies
\begin{equation}
\frac{\partial\Phi}{\partial\zeta}=2A(\zeta,\zeta^{+}),\quad\frac{\partial\Phi}{\partial\zeta^{+}}=2A^{\ast}(\zeta,\zeta^{+}).
\end{equation}
Thus the potential $\Phi(\zeta,\zeta^{+})$ is in general complex
for complex variables $(\zeta,\zeta^{+})$. In the situation where
$\zeta^{+}=\zeta^{*}$, it is directly checked that the potential
$\Phi(\zeta,\zeta^{+})$ is real. The stationary points can be calculated
by using the first derivatives, which are divided into three groups:
the origin solution ($\zeta=\zeta^{+}=0$), the classical solutions
($\zeta^{+}=\zeta^{*}$) and the nonclassical solutions ($\zeta^{+}=-\zeta^{*}$).
Here we are interested in the quantum tunneling between the classical
stationary points through the origin.

Given the classical stationary points $\zeta^{(c)}=re^{i\psi}$, we
will introduce the transformation 
\begin{eqnarray}
u & = & e^{-i\theta/2}e^{i\phi}\zeta+e^{i\theta/2}e^{-i\phi}\zeta^{+},\nonumber \\
v & = & e^{-i\theta/2}e^{-i\phi}\zeta-e^{i\theta/2}e^{i\phi}\zeta^{+},\label{eq:transform-2}
\end{eqnarray}
where $\phi=\psi-\theta/2$. Then the inverse transformation takes
the form 
\begin{eqnarray}
\zeta & = & e^{i\theta/2}\frac{e^{i\phi}u+e^{-i\phi}v}{\cos(2\phi)},\nonumber \\
\zeta^{+} & = & e^{-i\theta/2}\frac{e^{-i\phi}u-e^{i\phi}v}{\cos(2\phi)}.
\end{eqnarray}
In this case, the Fokker-Planck equation is transformed to 
\begin{eqnarray}
\frac{\partial P}{\partial\tau} & = & \left\{ \frac{\partial}{\partial u}\left[e^{i(\theta/2+\phi)}A(u,v)+h.c.\right]+\frac{\partial}{\partial v}\left[e^{i(\theta/2-\phi)}A(u,v)-h.c.\right]\right.\nonumber \\
 &  & \left.+\frac{\partial^{2}}{\partial u^{2}}\frac{\cos(2\phi)}{n}+\frac{\partial^{2}}{\partial v^{2}}\frac{\cos(2\phi)}{n}\right\} P.
\end{eqnarray}
The notation $h.c.$ indicates hermitian conjugate terms obtained
by the replacement of $u\rightarrow u$, $v\to-v$ and the conjugation
of all complex parameters. Considered the manifold~(\ref{eq:manifold-1}),
it is directly checked that $v$ is real on this manifold except on
the boundaries, while $u$ is in general complex. But for the situation
of $\zeta=\zeta^{+}$, we have $v=0$ with $u$ is real. All the points
on the line of $v=0$ with real $u$ have real potential $\Phi(u,v)$
as well as the real second derivatives $\Phi_{uu}$ and $\Phi_{vv}$,
which is proved in the \sref{subsec:Tunneling-rate}. The classical
stationary points and the origin solution are all located on the line
of $v=0$ with real $u$. We suppose that the classical stationary
points are local minima and the origin is the saddle point, as we
have in \sref{subsec:Local-stationary-points}. As discussed in
the Ref.~\cite{landauer1961frequency}, the quantum tunneling will
take place through the direction of $u$ because of the symmetry.
In the following, we will reproduce the analysis of \cite{landauer1961frequency}
so that the potential-barrier approximation can be generalized for
the complex potential cases.

As introduced in Ref.~\cite{landauer1961frequency}, the current
flow from a local minimum of the potential, located at negative $u$
and labelled as $C_{1}$, to the another minimum, located at positive
$u$ and labelled as $C_{2}$, has the form 
\begin{equation}
j=-\rho D\nabla\Phi-D\nabla\rho.\label{eq:flow}
\end{equation}
Here $\rho$ is the density  and $D$ is the diffusion coefficient.
We have taken the zero-temperature limit as we did in the main text.
The equilibrium case with Boltzmann distribution, 
\begin{equation}
\rho\propto\exp(-\Phi),
\end{equation}
leads to $j=0$. Thus we set 
\begin{equation}
\rho=\eta\exp(-\Phi),
\end{equation}
in the non-equilibrium case, then the variation of $\eta$ indicates
the extent of the deviation from equilibrium and the flow~(\ref{eq:flow})
becomes 
\begin{equation}
j=-D(\nabla\eta)\exp(-\Phi).\label{eq:flow_2}
\end{equation}
In the following, we assume that $D$ is constant in the neighborhood
of the saddle point, which is exactly true in our situation where
$D=\cos(2\phi)/n$. To obtain the magnitude of current~(\ref{eq:flow_2}),
this assumption requires a relatively large value of $\nabla\eta$
near the saddle point, where $\exp(-\Phi)$ is small, and much smaller
values of $\nabla\eta$ near the minima, where $\exp(\Phi)$ has larger
values. Therefore the major departures from equilibrium take place
only in the neighborhood of the saddle point.

For our case where the two potential minima are at the same value,
Ref.~\cite{landauer1961frequency} has shown that at the symmetry
plane, $j$ is perpendicular to the symmetry plane, and $j$ has the
same direction throughout the neighborhood of the saddle points. Here
we will use the same assumption where $u$ has been proved to be this
direction. Then equation~(\ref{eq:flow_2}) tells us that $\eta$
is only a function of $u$ in the neighborhood of the saddle point.
Therefore, we can integrate equation~(\ref{eq:flow_2}). 
\begin{equation}
\eta(u)=-\int_{0}^{u}\frac{j_{u}}{D}\exp[\Phi(u')]du'.\label{eq:eta}
\end{equation}
In the saddle-point neighborhood, the potential $\Phi$ depends quadratically
on the spatial coordinates 
\begin{eqnarray}
\Phi & = & \Phi^{(o)}+\frac{1}{2}\Phi_{uu}^{(o)}u^{2}+\frac{1}{2}\Phi_{vv}^{(o)}v^{2}\nonumber \\
 & = & \Phi^{(o)}-\frac{1}{2}\xi_{u}^{(o)}u^{2}+\frac{1}{2}\xi_{v}^{(o)}v^{2},
\end{eqnarray}
where the second derivatives $\xi_{u}^{(o)}=-\Phi_{uu}^{(o)}$, $\xi_{v}^{(o)}=\Phi_{vv}^{(o)}$
are both positive due to the manifold of the saddle-point neighborhood.
Thus we have 
\begin{eqnarray}
\eta(u) & = & -D^{-1}\int_{0}^{u}j_{u}\exp\left[\Phi^{(o)}-\frac{\xi_{u}^{(o)}(u')^{2}}{2}+\frac{\xi_{v}^{(o)}v^{2}}{2}\right]du'.
\end{eqnarray}
The continuity of current, $\nabla\cdot j=0$, requires that $j_{u}$
be independent of $u$. Considering that $\eta$ is only dependent
on $u$ in the neighborhood of the saddle point, the factor $j_{u}\exp[\xi_{v}^{(o)}v^{2}/2]$
is thus a constant. The only remaining variable is $\exp[-\xi_{u}^{(o)}(u')^{2}/2]$.
The integrand is then large only at the saddle point $u=0$, and diminishes
rapidly. Therefore, at a relatively short distance away from the saddle
point, $\eta(u)$ approaches a constant limiting value: a positive
value in the minimum $C_{1}$ and a negative value in the minimum
$C_{2}$.

Next, we will evaluate the population difference $\Delta$, which
is equal to twice the population of the classical minimum $C_{1}$,
\begin{eqnarray}
\Delta & = & 2\int_{C_{1}}dudv\,\eta(C_{1})\exp(-\Phi)\nonumber \\
 & = & 2\eta(C_{1})\exp\left(-\Phi^{(c)}\right)\int_{C_{1}}dudv\,\exp\left(-\frac{\xi_{u}^{(c)}u^{2}+\xi_{v}^{(c)}v^{2}}{2}\right).
\end{eqnarray}
Here we have used an expansion appropriate to the minimum $C_{1}$
\begin{eqnarray}
\Phi & = & \Phi^{(c)}+\frac{1}{2}\Phi_{uu}^{(c)}u^{2}+\frac{1}{2}\Phi_{vv}^{(c)}v^{2}\nonumber \\
 & = & \Phi^{(c)}+\frac{1}{2}\xi_{u}^{(c)}u^{2}+\frac{1}{2}\xi_{v}^{(c)}v^{2},
\end{eqnarray}
where $\xi_{u}^{(c)}=\Phi_{uu}^{(c)}$, $\xi_{v}^{(c)}=\Phi_{vv}^{(c)}$
are both positive second derivatives due to the manifold of the neighborhood
of the minimal point. Then we will get the difference 
\begin{equation}
\Delta=2\eta(C_{1})\exp\left(-\Phi^{(c)}\right)\left(\frac{2\pi}{\xi_{u}^{(c)}}\right)^{1/2}\left(\frac{2\pi}{\xi_{v}^{(c)}}\right)^{1/2}.
\end{equation}
In order to obtain the tunneling time, we need to evaluate the total
current $J$ crossing the saddle point as well. Since $\eta$ is only
dependent on $u$ in the neighborhood of the saddle point, the equation~(\ref{eq:flow_2})
is equivalent to 
\begin{equation}
j_{u}=-D\left(\frac{\partial\eta}{\partial u}\right)\exp\left(-\Phi\right).
\end{equation}
In the symmetry plane containing the saddle point ($u=0$), we will
then obtain 
\begin{equation}
j_{u}=-D\left(\frac{\partial\eta}{\partial u}\right)_{u=0}\exp\left(-\Phi^{(o)}-\frac{\xi_{v}^{(o)}v^{2}}{2}\right).
\end{equation}
Integrating over $v$ gives the total current 
\begin{eqnarray}
J & = & -D\left(\frac{\partial\eta}{\partial u}\right)_{u=0}\exp\left(-\Phi^{(o)}\right)\int dv\,\exp\left(-\frac{\xi_{v}^{(o)}v^{2}}{2}\right)\nonumber \\
 & = & -D\left(\frac{\partial\eta}{\partial u}\right)_{u=0}\exp\left(-\Phi^{(o)}\right)\left(\frac{2\pi}{\xi_{v}^{(o)}}\right)^{1/2}.
\end{eqnarray}
The dimensionless tunneling time is then obtained as 
\begin{eqnarray}
\tau_{tunnel} & = & \frac{\Delta}{J}\\
 & = & -\frac{2\sqrt{2\pi}}{D}\frac{\eta(C_{1})}{\left(\partial\eta/\partial u\right)_{u=0}}\left(\frac{\xi_{v}^{(o)}}{\xi_{u}^{(c)}\xi_{v}^{(c)}}\right)^{\frac{1}{2}}\exp\left(\Phi^{(o)}-\Phi^{(c)}\right).\nonumber 
\end{eqnarray}
Now we need to evaluate $\eta(C_{1})/(\partial\eta/\partial u)_{u=0}$.
From the equation~\eref{eq:eta}, we will get 
\begin{equation}
\eta(C_{1})=-\int_{0}^{C_{1}}\frac{j_{u}}{D}\exp\left(\Phi^{(o)}-\frac{\xi_{u}^{(o)}u^{2}}{2}\right)du,
\end{equation}
where $j_{u}$ is the current density at the saddle point, which takes
the form via equation~(\ref{eq:flow_2})
\begin{equation}
j_{u}^{(o)}=-D\left(\frac{\partial\eta}{\partial u}\right)_{u=0,v=0}\exp\left(-\Phi^{(o)}\right).
\end{equation}
Then we have 
\begin{eqnarray}
\eta(C_{1}) & = & \left(\frac{\partial\eta}{\partial u}\right)_{u=0,v=0}\int_{0}^{C_{1}}\exp\left(-\frac{\xi_{u}^{(o)}u^{2}}{2}\right)du\nonumber \\
 & \simeq & -\frac{1}{2}\left(\frac{\partial\eta}{\partial u}\right)_{u=0}\left(\frac{2\pi}{\xi_{u}^{(o)}}\right)^{1/2}.
\end{eqnarray}
Thus, we get 
\begin{equation}
\frac{\eta(C_{1})}{\left(\partial\eta/\partial u\right)_{u=0}}=-\frac{1}{2}\left(\frac{2\pi}{\xi_{u}^{(o)}}\right)^{1/2}.
\end{equation}
The tunneling time in dimensionless units then takes the form 
\begin{eqnarray}
\tau_{tunnel} & = & \frac{2\pi}{D}\left(\frac{\xi_{v}^{(o)}}{\xi_{u}^{(o)}\xi_{u}^{(c)}\xi_{v}^{(c)}}\right)^{1/2}\exp(\Phi^{(o)}-\Phi^{(c)})\nonumber \\
 & = & \frac{2\pi n}{\cos(2\phi)}\left[\frac{-\Phi_{vv}^{(o)}}{\Phi_{uu}^{(o)}\Phi_{uu}^{(c)}\Phi_{vv}^{(c)}}\right]^{\frac{1}{2}}\exp(\Phi^{(o)}-\Phi^{(c)}).
\end{eqnarray}
Noting that we have rescaled the time as $\tau=\mathcal{E}t$ in \sref{sec:Adiabatic-Elimination}
and $n=\mathcal{E}/|g|$, the dimensional tunneling time of our system
is therefore 
\begin{eqnarray}
T & = & \frac{2\pi}{|g|\cos(2\phi)}\left[\frac{-\Phi_{vv}^{(o)}}{\Phi_{uu}^{(o)}\Phi_{uu}^{(c)}\Phi_{vv}^{(c)}}\right]^{\frac{1}{2}}\exp(\Phi^{(o)}-\Phi^{(c)}),\label{eq:tunneling_complex}
\end{eqnarray}
which was used to obtain the analytic expression for the tunneling
time in~(\ref{tunneling_time}).

\section*{References}

\bibliography{TimeCrystalParamp1}
 
\end{document}